\documentclass[aps,pra,groupedaddress,twocolumn, floatfix]{revtex4-1}

\usepackage{graphicx}
\usepackage{amsmath}
\usepackage{mathtools}
\usepackage{upgreek}
\usepackage{braket}

\newcommand{\AR}{{\textrm{A}}}
\newcommand{\BR}{{\textrm{B}}}

\allowdisplaybreaks

\begin{document}

\title{Pulsed excitation of Rydberg-atom--pair states in an ultracold Cs gas}

\author{Heiner Sa{\ss}mannshausen, Fr\'ed\'eric Merkt, and Johannes Deiglmayr}

\affiliation{Laboratory of Physical Chemistry, ETH Zurich, Switzerland}
\email{jdeiglma@ethz.ch}
\begin{abstract}
Pulsed laser excitation of a dense ultracold Cs vapor has been used to study the pairwise interactions between Cs atoms excited to $n$p$_{3/2}$ Rydberg states of principal quantum numbers in the range $n=22-36$. Molecular resonances were observed that correspond to excitation of Rydberg-atom--pair states correlated not only to the $n$p$_{3/2}+n$p$_{3/2}$ dissociation asymptotes, but also to $n$s$_{1/2}+(n+1)$s$_{1/2}$, $n$s$_{1/2}+n'$f$_{j}$, and $(n-4)$f$_{j}+(n-3)$f$_{j}\ (j=5/2,7/2)$ dissociation asymptotes. These pair resonances are interpreted as arising from dipole-dipole, and higher-order long-range-interaction terms between the Rydberg atoms on the basis of i) their spectral positions, ii) their response to static and pulsed electric fields, and iii) millimeter-wave spectra between pair states correlated to different pair-dissociation asymptotes. The Rydberg-atom--pair states were found to spontaneously decay by Penning ionization and the dynamics of the ionization process were investigated during the first 15\,$\mu$s following initial photoexcitation. To interpret the experimental observations, a potential model was derived that is based on the numerical determination of the eigenvalues and eigenfunctions of the long-range interaction Hamiltonian. With this potential model, which does not include adjustable parameters, all experimental observations could be accounted for, and the results demonstrate that long-range-interaction models provide a global and accurate description of interactions in ultracold Rydberg gases and that they correctly account for, and enable the analysis of, phenomena as diverse as the formation of Rydberg macrodimers, Penning ionization in dense Rydberg gases, and  Rydberg-excitation blockade effects.

\end{abstract}

\maketitle

\section{Introduction}
Long-range interactions between atoms in high Rydberg states are the origin of many scientific applications  involving Rydberg excitation of ultracold atoms. The long-range van der Waals and resonant dipole-dipole interactions between high-$n$ Rydberg states lead to effects such as the Rydberg excitation blockade\,\cite{singer2004,tong2004,vogt2006,Urban2009} that might be utilized in realizing scalable quantum gates for quantum information processing\,\cite{lukin2001,moller2008,isenhower2010,saffman2010}, the evolution of ultracold Rydberg gases to ultracold plasmas\,\cite{li2004}, as well as strongly nonlinear optical processes \,\cite{Peyronel2012,Dudin2012}, e.g., to generate single photons. While many of the recent experiments with ultracold Rydberg atoms rely on the interactions between the excited atoms, few experiments directly probe the potential-energy structure of the interacting Rydberg gas.

Rydberg-Rydberg interactions were first observed as density-dependent line broadening in the excitation of Cs atoms in a beam\,\cite{raimond1981}. The effects of these interactions became easier to observe in experiments on ultracold atoms because these enable much longer interaction times under conditions where the interatomic potential energy dominates the dynamics of the system. Asymmetric line shapes and interaction-induced shifts have been observed, e.g., in millimeter-wave spectra of transitions between Rydberg states\,\cite{han2009}. Direct measurements of the long-range van der Waals and the resonant dipole-dipole interaction between two localized Rydberg atoms as a function of the interatomic separation were only reported recently\,\cite{Ravets2014,beguin2013}.

We present experiments in which Cs Rydberg-atom pairs are detected as sharp resonances in laser-excitation spectra of ultracold, dense samples of Cs. The resonances are assigned and quantitatively analyzed using a model describing the interactions between Cs Rydberg atoms using long-range molecular potential curves that do not involve any adjustable parameters.
Our experiments involve the pulsed excitation of pairs of interacting Cs Rydberg atoms with intense laser fields and enable the observation of molecular states connected to different pair-dissociation asymptotes. Whereas in our experiments only $n$p Rydberg states are accessible by selection rules for isolated atoms, Rydberg states can be observed as resonances near $n$s$_{1/2}(n+1)$s$_{1/2}$, $n$s$_{1/2}(n-3)$f$_{j}$, and $(n-4)$f$_{j}(n-3)$f$_{j}$ asymptotes as a result of dipole-dipole, dipole-quadrupole\,\cite{deiglmayr2014}, and higher-order terms of the multipole expansion series.  Our observations are related to earlier work describing similar phenomena in ultracold Rydberg gases of rubidium~\cite{farooqi2003,stanojevic2006,stanojevic2008,samboy2011} and cesium~\cite{overstreet2007,overstreet2009} at values of the principal quantum number $n$ around 70. Compared to these previous investigations of molecular resonances, we study these resonances, referred to as pair resonances hereafter, at lower principal quantum numbers in the range $22\leq n\leq36$. Additionally to pair resonances observed in the immediate vicinity of dissociation thresholds, we also observe
sharp spectral structures originating from local maxima in the molecular potentials and from avoided crossings between different potential curves for $n<27$.
The positions, widths and relative intensities of these resonances are sensitive to the details of the molecular potentials at long range. We exploit this sensitivity to validate our Rydberg-atom--pair interaction model, which we then use to describe the Penning ionization processes that take place in our dense Rydberg gas on timescales of several hundreds of nanoseconds, and the asymmetric broadening observed in millimeter-wave transitions between pairs of Rydberg atoms.

This paper is organized as follows: The experimental apparatus and the measurement techniques are described in Section II. The experimental results are presented in Section III.
The Rydberg-pair-interaction model we developed to account for the experimental observations is presented in Section IV. The main aspects of the paper are summarized in Section V and details concerning the calculations are presented in the appendix.

\section{Experimental}
All experiments are performed on ultracold samples of cesium atoms at a density of ${\sim}10^{12}\,\textrm{cm}^{-3}$, a particle number of ${\sim}10^6$, and a translational temperature of ${\sim}40$\,$\mu$K. The samples are released from a far-detuned crossed optical dipole trap ($\lambda=1064$~nm, $P=10$~Watt, ${\sim}80$\,$\mu$m 1/${\rm e}^2$ radius) which is loaded from a magneto-optical trap (MOT)\,\cite{sassmannshausen2013}. The transfer from the MOT to the dipole trap proceeds as follows: After a MOT-loading phase of ${\sim}80$\,ms, the fiber laser used for the dipole trap is turned on and, simultaneously, the magnetic-field gradient is ramped up, compressing the MOT for 10\,ms. The magnetic field is then turned off and the atoms are cooled in an optical-molasses phase for 5\,ms. The MOT lasers are then turned off and the atoms are optically pumped into the lower hyperfine component of the 6s$_{1/2}$ electronic ground state. The atoms are held in the dipole trap for 10\, ms, during which time the noncaptured atoms can leave the photoexcitation region.
 The power of the optical-dipole-trap laser is reduced to its minimum value of 300\,mW after the cesium atoms in the optical dipole trap have been selectively pumped optically into the upper hyperfine state of the 6s$_{1/2}$ ground state, from where
they are photoexcited in single-photon transitions to $n$p$_{3/2}$ Rydberg states. The experimental cycle is repeated at a rate of 10 Hz.

The excitation laser is a frequency-doubled pulse-amplified ring dye laser (Coherent 899-21) delivering 4.4-ns-long Fourier-transform-limited UV pulses (${\sim}319$\,nm, 140\,MHz bandwidth) with pulse energies of up to 100\,$\mu$J. The laser is focused to a spot size of ${\sim}150$\,$\mu$m (1/e$^2$ radius) resulting in peak irradiances of up to 100\,MW/cm$^2$. The laser frequency is calibrated with a wavemeter (High Finesse WS6-200).
We have also employed single-color two-photon excitation of $n$s$_{1/2} \leftarrow 6$s$_{1/2}$ transitions using the pulse-amplified fundamental output of the ring laser at a wavelength of ${\sim}639$\,nm and a pulse energy of ${\sim}0.5$\,mJ.

Transitions are detected by monitoring the Cs$^+$ ion signal using a microchannel-plate (MCP) detector after state-selective pulsed-field ionization (PFI). For PFI, we employ a pulsed electric field with a rise time of 1\,$\mu$s and maximal field strength $F_{\rm max}=1.25$\,kV/cm.  Only $n$p Rydberg states with $n \geq 26$ can be field ionized under these conditions. We also observe the formation of ions on molecular resonances for Rydberg states with $n<26$, but only after introducing some delay between laser and electric-field pulse. After PFI and before the next MOT-loading phase, the atom cloud is characterized by saturated-absorption imaging. From the images, values for the particle numbers and densities are derived. On the atomic resonances the trap is fully depleted and the resulting ion signals strongly saturate our detection system, which is designed to be sensitive to single ions. The spectral positions of the atomic transitions are obtained from trap-loss spectra based on the saturated-absorption images.

 Millimeter-wave transitions between Rydberg states excited on atomic or pair resonances are measured using a millimeter-wave source consisting of a swept frequency synthesizer (Wiltron, Model 6769B) connected to an amplifier-multiplier chain (Virginia Diodes, WR6.5AMC, tunable range 110\,-\,170\,GHz). Typical interaction times with the millimeter-wave radiation are $1$\,$\mu$s and the transitions are detected using state-selective PFI and separating initial and final Rydberg states in the Cs$^+$-ion time-of-flight (TOF) traces.

Regular measurement and compensation of the stray electric and magnetic fields during the data acquisition period ensured that the magnitude of the stray electric and magnetic fields never exceeded 20\,mV/cm and 20\,mG, respectively, which is sufficiently low that these fields do not influence the behavior of the Rydberg states we study. The electric field was compensated by measuring and minimizing the quadratic Stark effect of high $n$p$_{3/2}$ Rydberg states \cite{sassmannshausen2013} and the magnetic field was compensated by measuring the Zeeman splitting of the $F'=4 \leftarrow F=3$ clock transition of the 6s$_{1/2}$ state using microwave spectroscopy.

\section{Experimental results and discussion}
\subsection{Excitation of dipole-coupled pair states}\label{sec:expExtPairstates}
The experimental spectra presented in this section were recorded by monitoring the Cs$^+$-ion yield resulting from the spontaneous ionization of the Rydberg atoms taking place in the first 5\,$\mu$s following photoexcitation in the dense ultracold sample as a function of the UV-laser frequency. Because of a Cooper minimum in the photoexcitation cross section from the 6s$_{1/2}$ ground state of Cs to the $n$p$_{1/2}$ ($\epsilon $p$_{1/2}$) states (continuum) with center located just above the ionization threshold~\cite{raimond1978}, transitions to $n$p$_{1/2}$ Rydberg states are more than 500 times weaker than to the corresponding $n$p$_{3/2}$ Rydberg states. Consequently,
the experiments focus on the $n$p$_{3/2}$ states. To facilitate the comparison between spectra recorded in the vicinity of different $n$p$_{3/2} \leftarrow 6$s$_{1/2}$ transitions, the origins of the frequency scale of the different spectra presented in the following are always set at the positions of the atomic resonances.

\begin{figure}
\begin{center}
\includegraphics[width=0.99\linewidth]{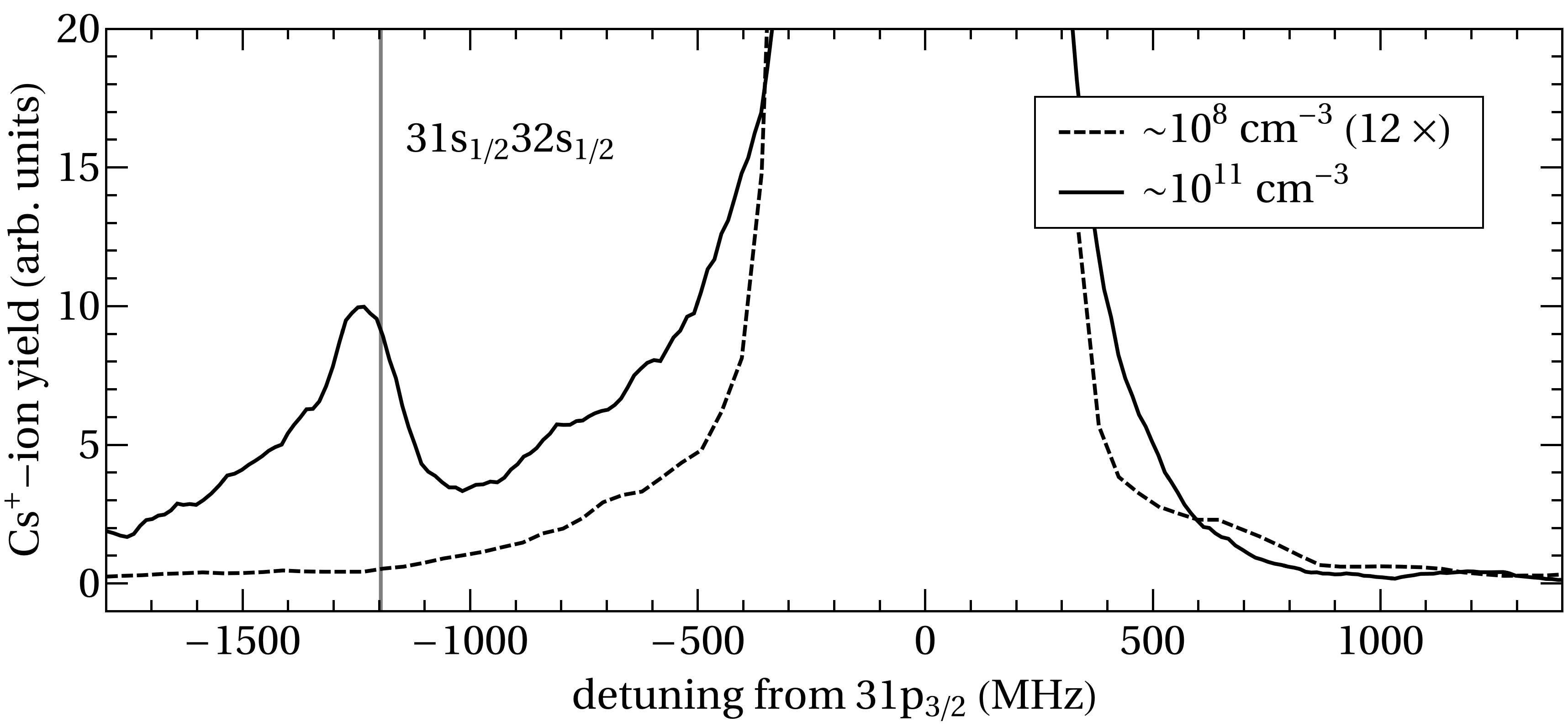}
\caption{UV-laser spectra near the 31p$_{3/2}\leftarrow6$s$_{1/2}$ transition of Cs at ground-state-atom densities of around $10^8$~atoms/cm$^3$ (dashed trace) and of $5 \cdot 10^{11}$~atoms/cm$^3$ (solid trace).}
\label{fig:density}
\end{center}
\end{figure}

Fig.\,\ref{fig:density} compares two spectra of the 31p$_{3/2} \leftarrow 6$s$_{1/2}$ transition recorded at a ground-state--atom density of ${\sim}10^8$\,atoms/cm$^3$ (dashed line, sample obtained by releasing the atoms from the MOT and letting them expand) and of ${\sim} 5 \cdot 10^{11}$\,atoms/cm$^3$ (full line, sample obtained after release from the optical dipole trap).  In both spectra, the atomic 31p$_{3/2} \leftarrow 6$s$_{1/2}$ transition is strongly saturated. The spectrum measured at lower atom density has been scaled up to match the wings of the 31p$_{3/2} \leftarrow 6$s$_{1/2}$ transition recorded at higher atom density. In the latter spectrum, an additional resonance is observed at a detuning of $-1200$\,MHz. The spectral position of this resonance cannot be assigned to any isolated-atom transition in Cs but matches closely the asymptotic energy of the 31s$_{1/2}32$s$_{1/2}$ pair state if one assumes a two-photon excitation, for which the actual detuning is $2 \times (-1200$\,MHz). The resonance has a red-degraded asymmetric line shape and a full width at half maximum (FWHM) of $2\times250$\,MHz.

\begin{figure}
\begin{center}
\includegraphics[width=0.99\linewidth]{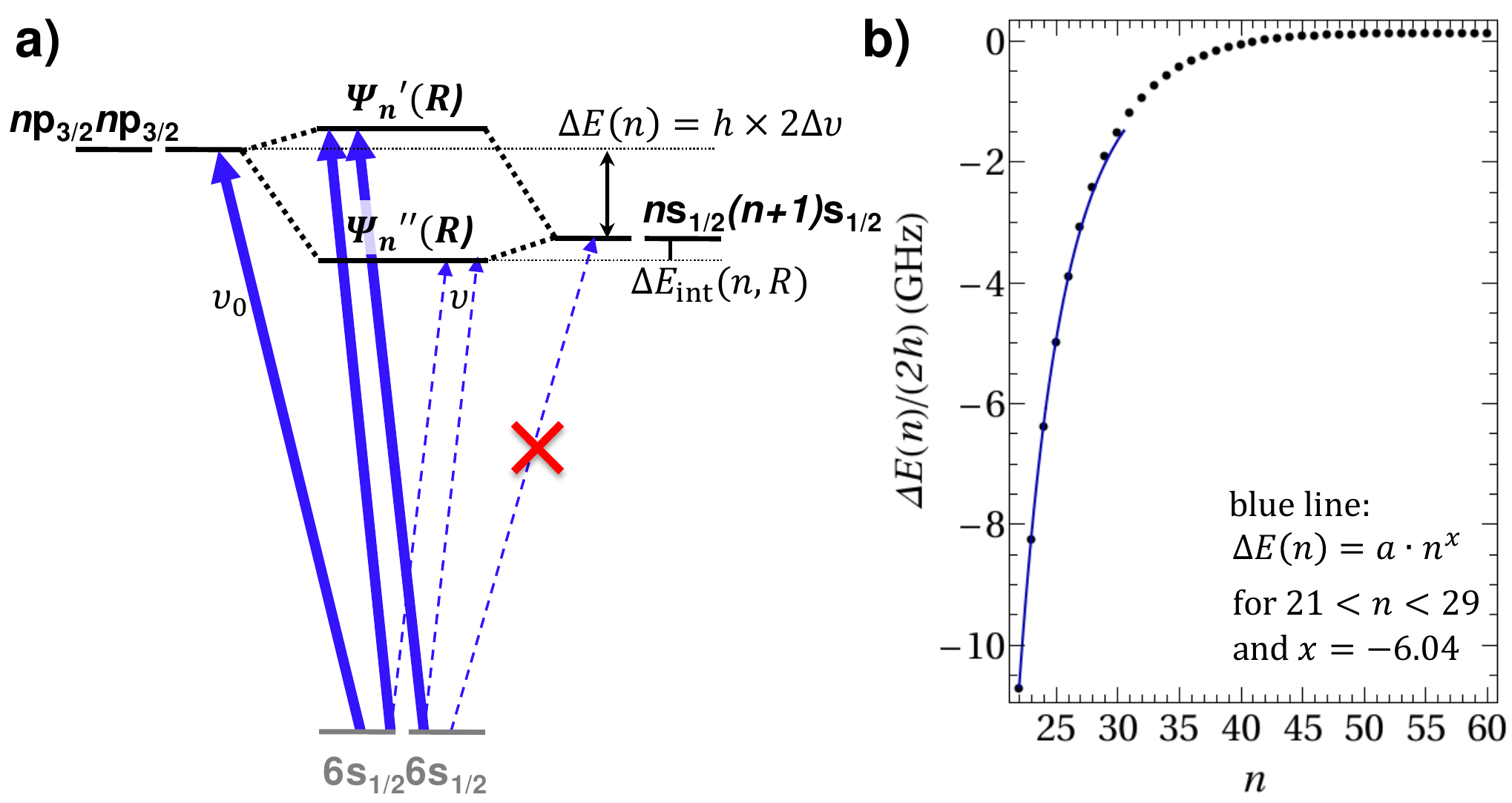}
\caption{(Color online) a) Simplified and schematic diagram of the interaction between the $n$s$_{1/2}(n+1)$s$_{1/2}$ and the $n$p$_{3/2}n$p$_{3/2}$ Rydberg-atom--pair states. The blue thick and the thin, dashed lines indicate strong and weak transitions, respectively, from the $6$s$_{1/2}$ ground state. Two-photon transitions are indicated by two parallel arrows. b) Energetic separation between the two pair states as a function of the principal quantum number $n$. The blue curve corresponds to the function $an^{-6.04}$ obtained by fitting the exponent to the energies of the dissociation asymptotes of $n$s$_{1/2}(n+1)$s$_{1/2}$ pair states relative to the $n$p$_{3/2}n$p$_{3/2}$ asymptotes.}
\label{fig:mo}
\end{center}
\end{figure}

In the absence of interactions between the Rydberg atoms, the single-color two-photon excitation of the 31s$_{1/2}32$s$_{1/2} \leftarrow 6$s$_{1/2}6$s$_{1/2}$ transition is prohibited because pure $s' \leftarrow s$ transitions are forbidden by the Laporte rule. However, under our experimental conditions, long-range interactions between the Rydberg atoms can lead to state mixing, especially between energetically close-lying dipole-dipole-coupled pair states such as the $n$s$_{1/2}(n+1)$s$_{1/2}$ and the $n$p$_{3/2}n$p$_{3/2}$ states, as schematically depicted in Fig.\,\ref{fig:mo}a. The molecular shifts $\Delta E_{\rm int}$ induced by the long-range interactions are expected to be small and resonances occur close to the $n$s$_{1/2}(n+1)$s$_{1/2}$ asymptotes. The $n$ dependence of the pair-state asymptote positions with respect to the atomic $n$p$_{3/2}$ resonance positions can be calculated as
\begin{equation}
\Delta \nu(n) = \frac{\Delta E(n)}{2h}=\frac{2 E_{n{\rm p}_{3/2}}-E_{n{\rm s}_{1/2}}-E_{(n+1){\rm s}_{1/2}}}{2h}
\label{eq:energiesI}
\end{equation}
with
\begin{equation}
E_{nl_j} = \frac{hR_{\rm Cs}}{(n-\delta_{l_j}(n))^2}.
\label{eq:energies}
\end{equation}
In Equation\,(\ref{eq:energies}), $R_{\rm Cs}= 3.289828299(20)\cdot 10^9$\,MHz is the Rydberg constant of Cs and $\delta_{l_j}(n)$ are the $l$- and weakly $j$- and $n$-dependent quantum defects. The $n$ dependence of the quantum defects can be expressed with the Ritz formula\,\cite{ritz1908}
\begin{align}
\delta_{l_j}(n)=\delta_{l_j}(\infty)-\frac{a_{l_j}}{(n-\delta_{l_j}(\infty))^2}.
\label{eq:ritz}
\end{align}
The parameters $\delta_{l_j}(\infty)$ and $a_{l_j}$ are taken from the literature\,\cite{goy1982}. The detuning $\Delta \nu(n)$ of the pair asymptotes decreases rapidly in the range from $n=20$ to $n=42$, where the detuning changes sign. In the range of $n$ values investigated, $\Delta \nu(n)$ approximately scales as $n^{-6}$ (see fit shown as blue line in Fig.\,\ref{fig:mo}b).
Because the frequency required for the two-photon excitation of the pair resonance is detuned from the atomic $n$p$_{3/2} \leftarrow 6$s$_{1/2}$ resonance, the two-photon resonance can only be observed at the high laser power achievable with a pulse-amplified laser system. It is also necessary to have a high ground-state-atom density to have a sufficient number of atom pairs at distances where the dipole-dipole interaction is strong enough to induce the state mixing illustrated in a simplified 2-level model in Fig.\,\ref{fig:mo}a). No pair resonances could be observed in spectra recorded with a 100\,mW continuous-wave UV laser under otherwise similar experimental conditions\,\cite{sassmannshausen2015}.

\begin{figure}
\begin{center}
\includegraphics[width=0.85\linewidth]{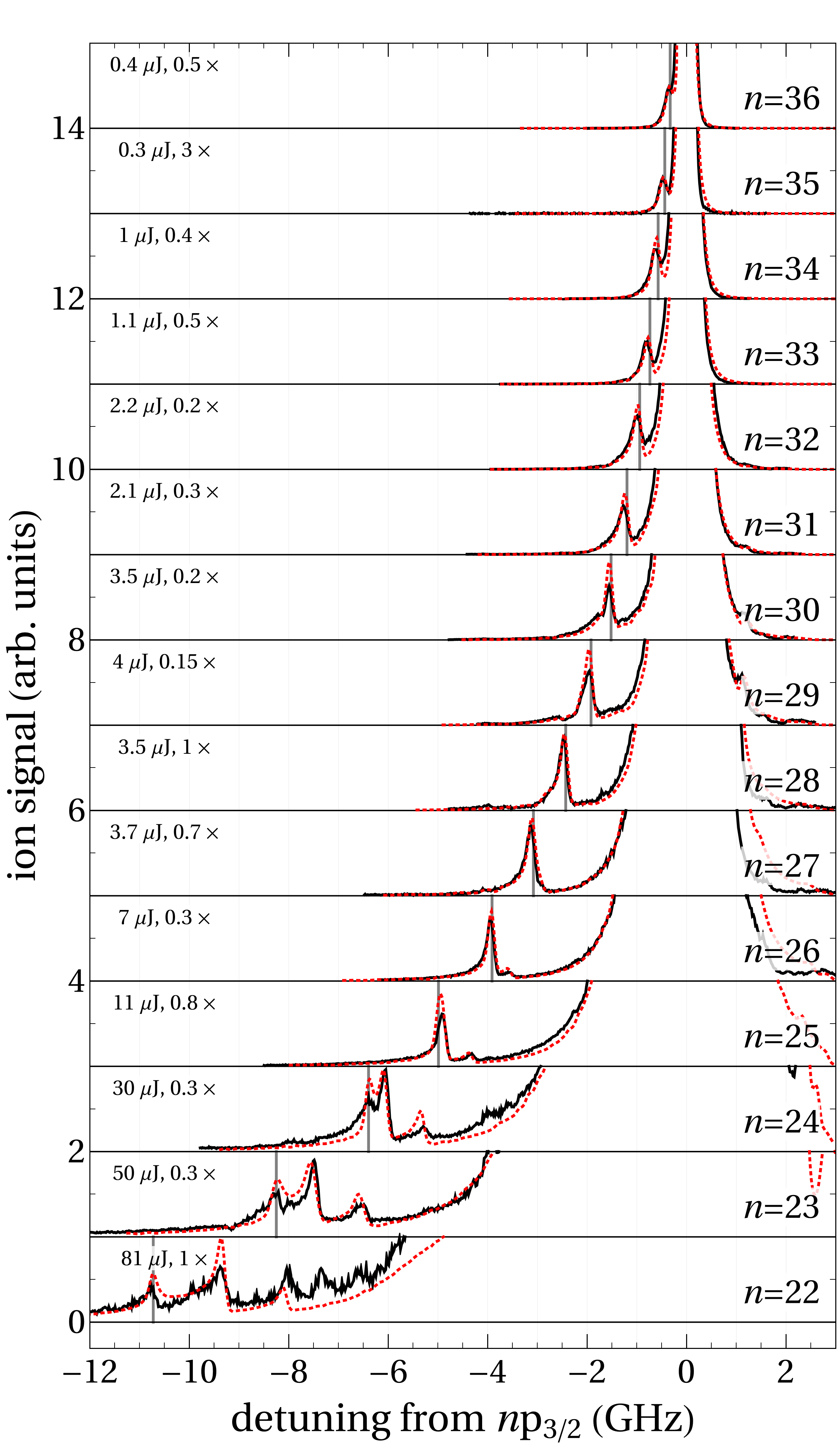}
\caption{(Color online) Pulsed-UV-laser spectra of transitions from the 6s$_{1/2}$ ground state of Cs  to $n$p$_{3/2}$ Rydberg states with $n$ between 22 and 36 (solid black traces). The Cs$^+$-ion signal is plotted as a function of the detuning from the atomic $n$p$_{3/2}\leftarrow 6$s$_{1/2}$ transition. The spectral positions corresponding to the asymptotic energy of $n$s$_{1/2}(n+1)$s$_{1/2}$ pair states calculated from experimental quantum defects are indicated by grey vertical lines. Simulated spectra are shown as red dashed lines. The pulse energies used to record the different experimental spectra and the relative intensity scaling factors of the experimental spectra are given on the left-hand side of each panel.}
\label{fig:allspectra}
\end{center}
\end{figure}

Rydberg-excitation spectra in the vicinity of $n$p$_{3/2} \leftarrow 6s_{1/2}$ transitions for $n=22-36$ are displayed in Fig.\,\ref{fig:allspectra}. At each $n$ value, we observe resonances on the low-frequency side of the atomic transitions that can be attributed to the two-photon excitation of two atoms to interacting pair states correlated to the $n$s$_{1/2}(n+1)$s$_{1/2}$ asymptotes. The positions of these asymptotes calculated with Eq.\,(\ref{eq:energies}) are marked by vertical gray lines in Fig.\,\ref{fig:allspectra}.

\begin{figure}
\begin{center}
\includegraphics[width=0.9\linewidth]{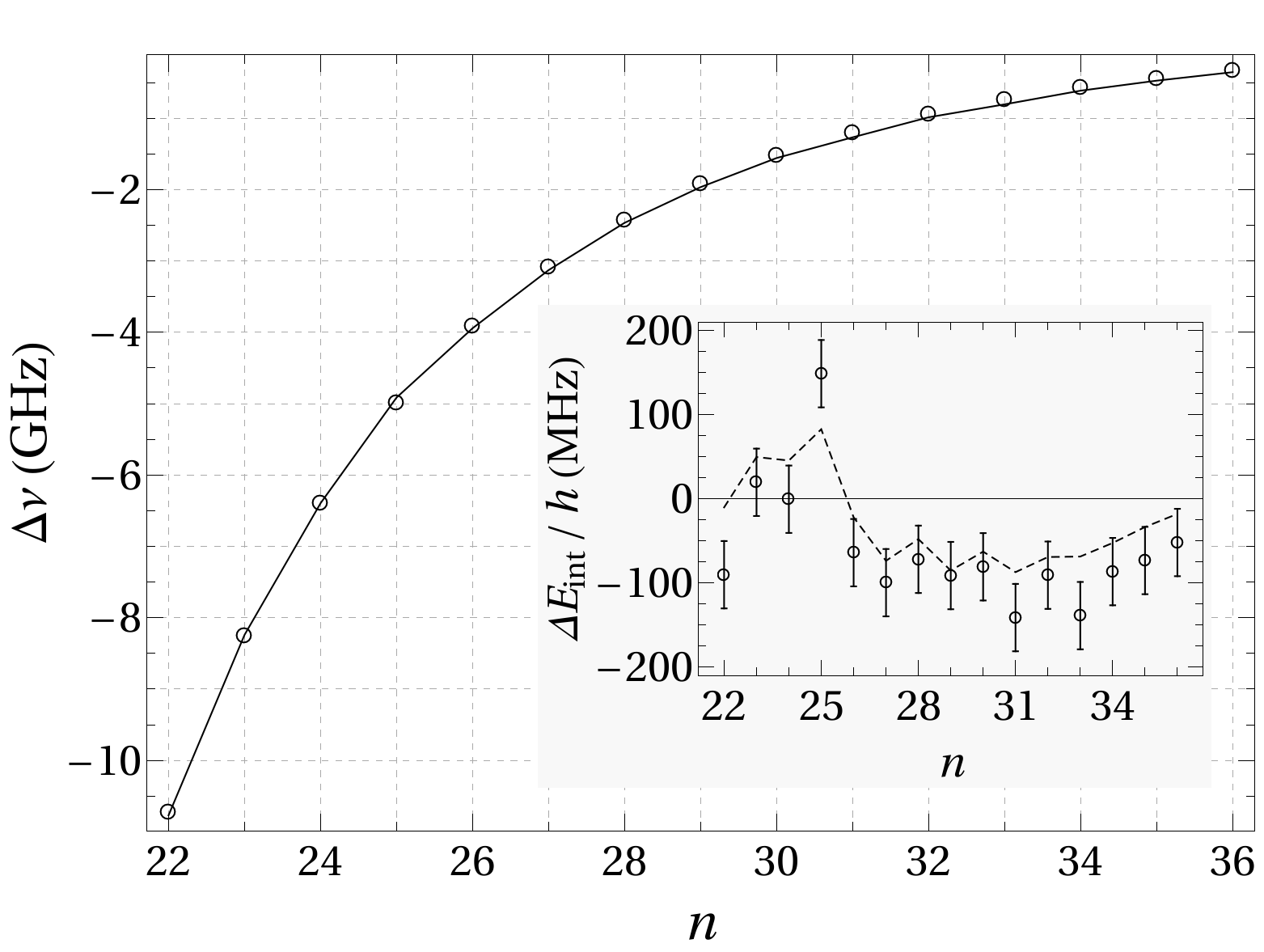}
\caption{Spectral positions of the $n$s$_{1/2}(n+1)$s$_{1/2}$ pair resonances relative to the atomic $n$p$_{3/2}$ states for values of the principal quantum number $n$ in the range between 22 and 36. The open circles represent the peak positions extracted from measured spectra and the black line corresponds to the relative positions of the asymptotes calculated from experimental values of the quantum defects of Cs using Eq.~\eqref{eq:energiesI}. The experimental (simulated) interaction-induced shifts $\Delta E_{\rm int}$ relative to the $n$s$_{1/2}(n+1)$s$_{1/2}$ pair-dissociation asymptotes are given in the inset as open circles (dashed line).}
\label{fig:ndependence}
\end{center}
\end{figure}

For all $n$ values for which measurements could be performed, i.e., for $n=22-36$, pair resonances are observed within $2 \times 75$\,MHz of the positions of the $n$s$_{1/2}(n+1)$s$_{1/2}$ asymptotes. The deviations are displayed on a magnified scale in the inset of Fig.\,\ref{fig:ndependence} (open circles). In most cases, the deviations are negative, which indicates that the pair resonances are typically located on the low-frequency side of the asymptotes, which is characteristic of attractive molecular potentials. The measurement at $n=25$ represents an exception and reveals a pair resonance located $2 \times 75$\,MHz above the 25s$_{1/2}26$s$_{1/2}$ asymptote. These shifts from the asymptotes, though small, are real and will be discussed in more detail in Section\,IV.

The increasing detuning of the pair resonances from the atomic transitions (Fig.\,\ref{fig:mo}) and the decreasing interaction strength with decreasing $n$ values lead to significantly smaller pair-state signals at lower $n$ values. This effect is compensated by using higher pulse energies in the excitation, as indicated on the left-hand side of Fig.\,\ref{fig:allspectra}. At the lowest $n$ values for which the Rydberg-atom--pair resonances could be observed ($n=22-26$ in Fig.\,\ref{fig:allspectra}), we also observe additional structures on their high-frequency side. These structures are discussed in more detail in Section IV. The dipole-quadrupole pair resonances presented in Ref. \cite{deiglmayr2014} are located at smaller detunings and are off the vertical scale of the panels of Fig.\,\ref{fig:allspectra}.

The close agreement between the positions of the $n$s$_{1/2}(n+1)$s$_{1/2}$ asymptotes and the molecular resonances near the $n$p$_{3/2} \leftarrow 6$s$_{1/2}$ atomic transitions in high-density samples (see Figs.\,\ref{fig:density} and \ref{fig:allspectra}) strongly suggests that the mechanism responsible for their observation is the dipole-dipole-coupling mechanism illustrated schematically in Fig.\,\ref{fig:mo}a). Further evidence for this mechanism was obtained from the measurement of the Stark effect and of millimeter-wave transitions discussed in the next subsections.

\subsection{Stark effect of the $n$s$_{1/2}(n+1)$s$_{1/2}$ and $n$p$_{3/2}n$p$_{3/2}$ pair states}
The atomic s and p Rydberg states of Cs exhibit quadratic Stark shifts in small external electric fields because these states have large quantum defects and are well separated from other Rydberg states of opposite parity. Therefore, when a small external electric field of strength $F$ is applied, one expects the positions of the $n$p$_{3/2} \leftarrow 6$s$_{1/2}$ transitions (and hence the positions of the $n$p$_{3/2}n$p$_{3/2}$ asymptotes) to be shifted by $\frac{1}{2}\alpha_{\rm p_{3/2}}(n)F^2$ and the positions of the $n$s$_{1/2}(n+1)$s$_{1/2}$ asymptotes to be shifted by $\frac{1}{2}F^2\left[\frac{1}{2}\alpha_{\rm s_{1/2}}(n)+\frac{1}{2}\alpha_{\rm s_{1/2}}(n+1)\right]$ in UV spectra of the type depicted in Figs.\,\ref{fig:density} and\,\ref{fig:allspectra}. The polarizabilities $\alpha_{l_{j}}$ are negative and much larger for p$_{3/2}$ Rydberg states than for s$_{1/2}$ Rydberg states of the same $n$ value. Consequently, a measurement of the Stark shifts of the pair resonances at low fields, for which the Stark shift is quadratic, enables one to distinguish between pair states of $n$p$_{3/2}n$p$_{3/2}$ and $n$s$_{1/2}(n+1)$s$_{1/2}$ character. The different polarizabilities of these two pair states can also be used to bring them into resonance with electric fields and to observe effects of resonant dipole-dipole interactions, which scale as $n^4/R^3$\,\cite{Westermann2006,vogt2006,gurian2012}. For the measurements relevant for this investigation, the dipole-dipole interaction between $n$p$_{3/2}n$p$_{3/2}$ and $n$s$_{1/2}(n+1)$s$_{1/2}$ pair states is nonresonant and of shorter range (scaling with $1/R^6$) so that the measurement of the pair-state Stark shifts necessitates high laser intensities and high atom densities.

\begin{figure}
\begin{center}
\includegraphics[width=0.95\linewidth]{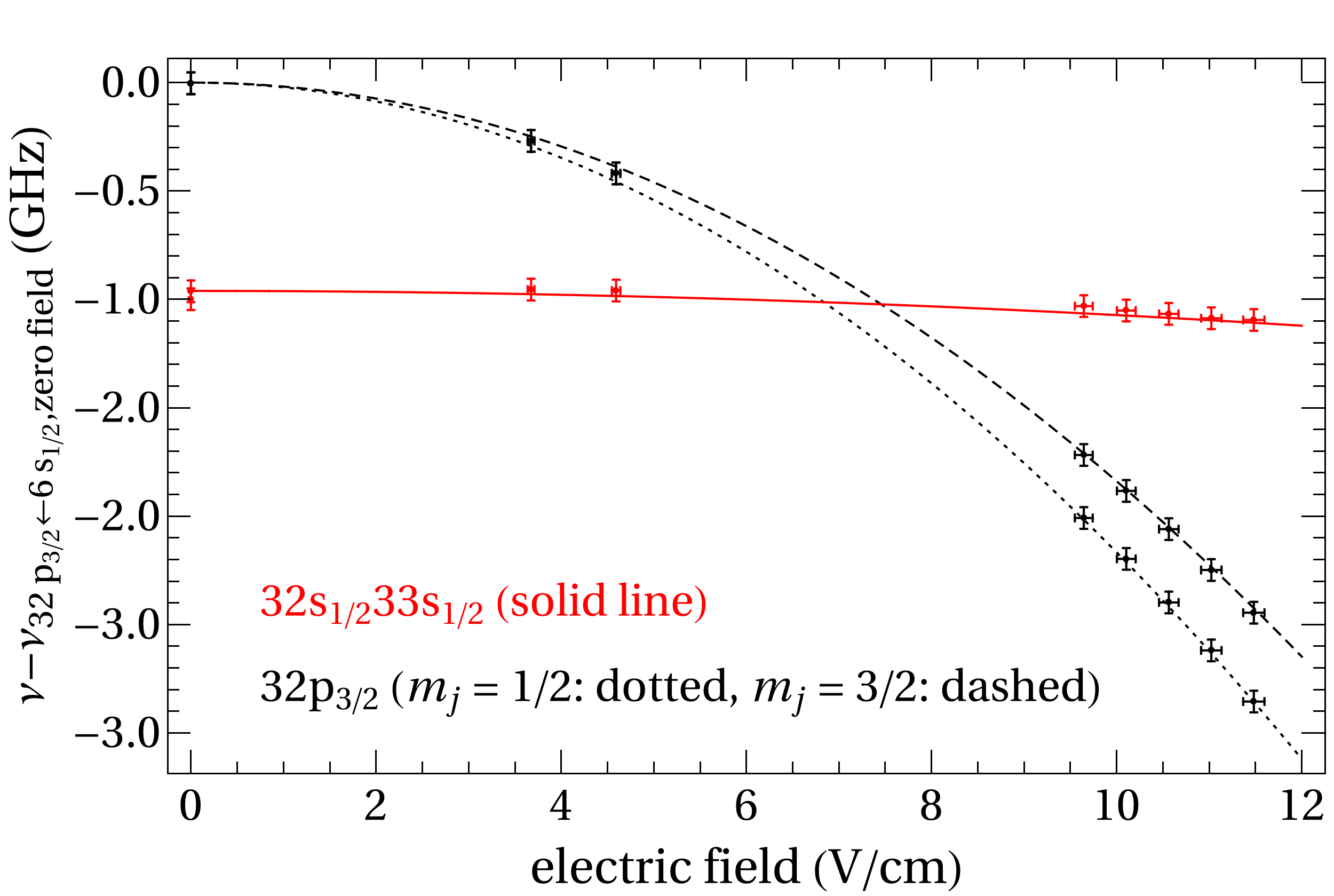}
\caption{(Color online) Stark shifts of the atomic 32p$_{3/2}\leftarrow 6$s$_{1/2}$ transition of Cs (black points) and of the 32s$_{1/2}33$s$_{1/2}\leftarrow 6$s$_{1/2}6$s$_{1/2}$ two-photon pair resonance (red points). The calculated quadratic Stark shifts of the atomic and the pair (averaged atomic Stark shifts of the two separate Rydberg atoms) resonances are given as dashed and dotted black and solid red lines, respectively.}
\label{fig:foerster}
\end{center}
\end{figure}

We have measured Rydberg excitation spectra of Cs in the vicinity of the 32p$_{3/2} \leftarrow 6$s$_{1/2}$ transition in weak external electric fields between 0\,V/cm and 12\,V/cm. The electric-field strength was varied by applying different electric potentials to sets of electrodes surrounding the ultracold atoms, as described in Ref.\,\cite{sassmannshausen2013}. The measured spectra have the same general appearance as the spectra depicted in Fig.\,\ref{fig:allspectra}, but the strong atomic and weak pair resonances are shifted by the quadratic Stark effect, and the atomic resonance splits into the two magnetic components with $|m_j|=1/2$ and $3/2$. All measured positions with respect to the field-free 32p$_{3/2} \leftarrow 6$s$_{1/2}$ atomic transition frequency are plotted as a function of the applied electric field in Fig.\,\ref{fig:foerster}, where the black and red dots correspond to the atomic and pair resonances, respectively. The observed quadratic Stark shift and splitting of the 32p$_{3/2}$ Rydberg state was used to calibrate the applied electric field by comparison to the calculated Stark shifts of the $|m_j|=3/2$ (dashed line) and $|m_j|=1/2$ (dotted line) components of the 32p$_{3/2}$ state. The Stark shift of the 32s$_{1/2}33$s$_{1/2}$ pair state is much smaller and matches the average of the calculated Stark shifts of two noninteracting atoms in the 32s$_{1/2}$ and 33s$_{1/2}$ states (red line in Fig.\,\ref{fig:foerster}) within the experimental uncertainty. The 32p$_{3/2}32$p$_{3/2}$ and the 32s$_{1/2}33$s$_{1/2}$ asymptotic energies are degenerate at an electric field of $\approx7$\,V/cm. At higher electric-field strengths, we observe the 32s$_{1/2}33$s$_{1/2}$ resonance on the high-frequency side of the transitions to the 32p$_{3/2}|m_j|=1/2,3/2$ states. The agreement between measured and calculated Stark shifts indicates that the Rydberg atoms excited at the positions of the pair resonances have predominantly s$_{1/2}$ character and that the pairs are excited close to their asymptote, at long range.

\subsection{Millimeter-wave spectroscopy of transitions between Rydberg-atom--pair states}

\begin{figure}
\begin{center}
\includegraphics[width=0.99\linewidth]{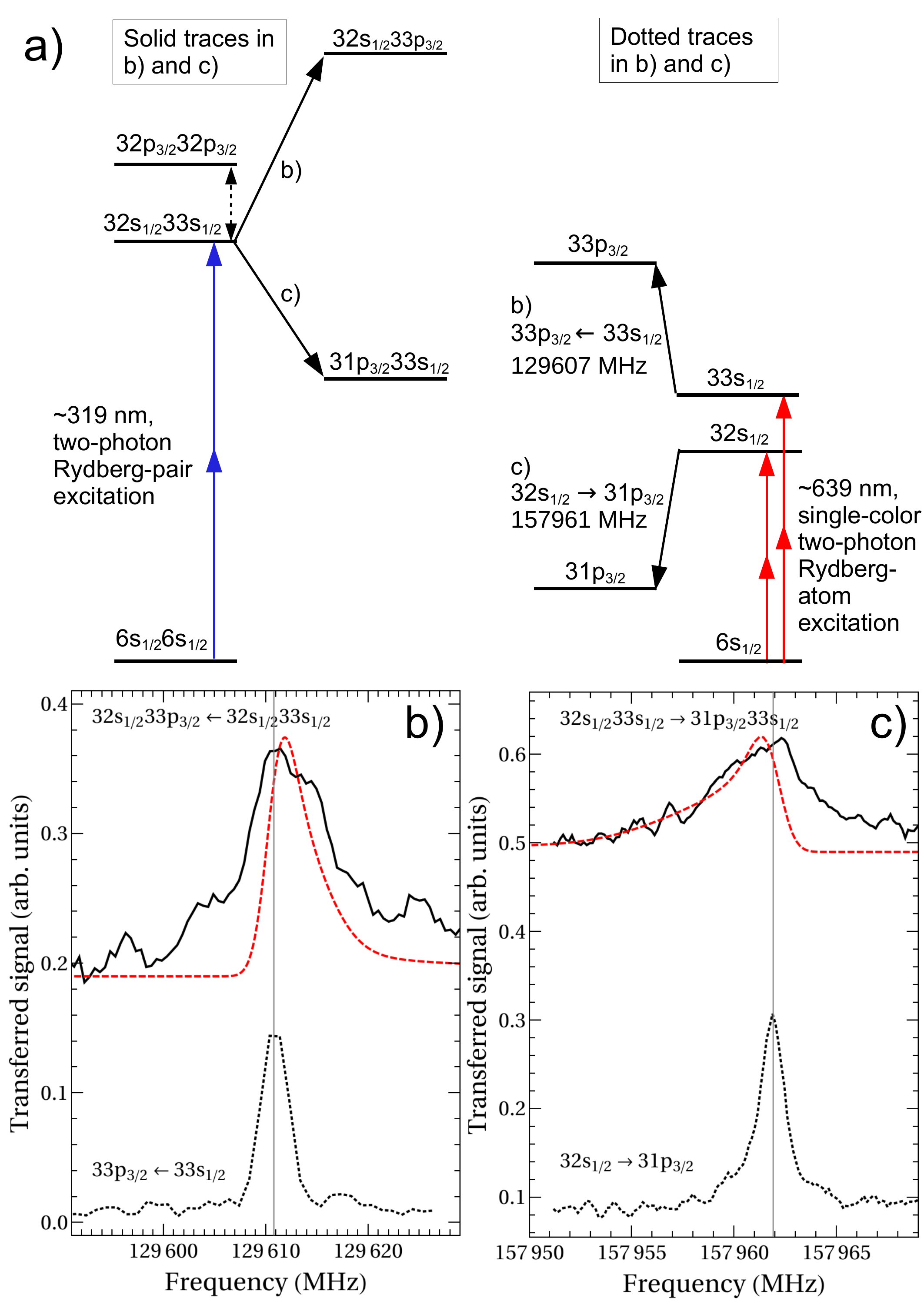}
\caption{(Color online) a) Schematic energy diagram of the $32$s$_{1/2}33$s$_{1/2} \leftarrow 6$s$_{1/2}6$s$_{1/2}$ pair state excitation (left-hand side, blue arrows) and the single-color two-photon excitation of $n$s$_{1/2} \leftarrow 6$s$_{1/2}$ transitions (right-hand side, red arrows). Millimeter-wave transitions are indicated with black arrows in the energy-level diagrams. Experimental millimeter-wave spectra of the 33p$_{3/2} \leftarrow 33$s$_{1/2}$ (b) and 32s$_{1/2} \rightarrow  31$p$_{3/2}$ (c) transitions recorded following excitation of the 32s$_{1/2}33$s$_{1/2}$ pair state (solid black traces) and following excitation of the atomic  $n$s$_{1/2} \leftarrow 6$s$_{1/2}$ transitions (dashed black traces), respectively.
The spectra represent the ratio of ions detected in the integration window of the relevant final Rydberg states over the ions detected in the integration windows corresponding to both the final and initial Rydberg states (see Fig.\,\ref{fig:tofs}).
The red dashed lines are spectra simulated with a model taking into account all relevant Rydberg-Rydberg--pair potentials and considering the Penning-ionization dynamics, as explained in more detail in the text (see Section IV D).}
\label{fig:mmspecs}
\end{center}
\end{figure}

\begin{figure}
\begin{center}
\includegraphics[width=0.8\linewidth]{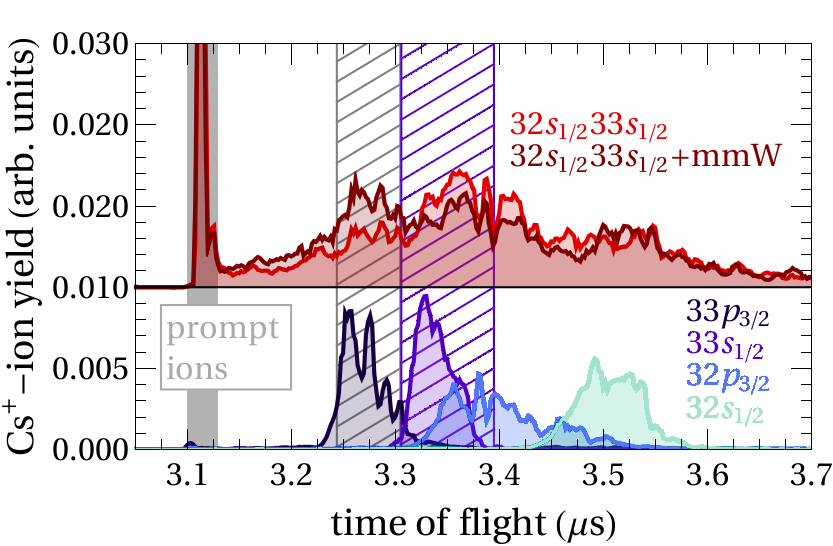}
\caption{(Color online) Cs$^+$-ion time-of-flight (TOF) distributions measured using a pulsed extraction field of 1.25\,kV/cm with a 1\,$\mu$s rise time. Top panel: TOF distributions following the two-photon excitation of the 32s$_{1/2}$33s$_{1/2}$ pair state without (red trace) and with millimeter-wave radiation (mmW) with the frequency tuned to the 33p$_{3/2} \leftarrow 33$s$_{1/2}$  transition (brown trace). Bottom panel: TOF distributions following the one-photon excitation of the atomic 33p$_{3/2}$ (black trace) and 32p$_{3/2}$ (blue trace) states, and the two-photon excitation of the atomic 33s$_{1/2}$ (violet trace) and 32s$_{1/2}$ (turquois trace) states. The $n$s$_{1/2}$ states have been excited from the 6s$_{1/2}$ ground state in single-color two-photon transitions. Hatched areas indicate the integration windows used to extract the spectra shown in Fig.\,\ref{fig:mmspecs}b).}
\label{fig:tofs}
\end{center}
\end{figure}

The ability to observe well-resolved transitions to $n$s$_{1/2}(n+1)$s$_{1/2}$ dipole-dipole-coupled pair states with good signal-to-noise ratio offers the possibility to further characterize these states by studying transitions to neighboring dipole-dipole-coupled states using millimeter-wave radiation. In a similar fashion microwave transitions between pairs of Rb Rydberg atoms were studied by Y. Yu \textit{et al.}~\cite{yu_2013}.

At large internuclear distances, one expects the most intense transitions to be those to pair states located near $n$s$_{1/2}n_1$p$_{j}$ and $n_2$p$_{j}(n+1)$s$_{1/2}$ asymptotes, where the values of the principal quantum numbers $n_1$ and $n_2$ are close to the value of $n$. Our investigation focused on the 32s$_{1/2}33$s$_{1/2}$ pair state and on the millimeter-wave transitions to the pair states near the 32s$_{1/2}33$p$_{3/2}$ and 31p$_{3/2}33$s$_{1/2}$ asymptotes, which lie 129611\,MHz above, and 157962\,MHz below the 32s$_{1/2}33$s$_{1/2}$ asymptote, respectively (see Fig.\,\ref{fig:mmspecs}a)). The detection of the millimeter-wave transitions was achieved by delayed pulsed-field ionization with a slowly rising electric-field ramp (rise time 1\,$\mu$s) and monitoring the Cs$^+$-ion time-of-flight (TOF) spectra. To calibrate the TOF spectra, experiments were first performed in low-density samples following single-photon laser excitation to the $n$p$_{3/2}$ ($n=31,33$) and nonresonant single-color two-photon excitation to the $n^\prime$s$_{1/2}$ ($n^\prime=32,33$) Rydberg states of Cs. The resulting Cs$^+$ TOF traces are displayed in the lower panel of Fig.\,\ref{fig:tofs} and reveal distinct maxima for the four Cs Rydberg states. This state selectivity makes it possible to record millimeter-wave spectra of the 33p$_{3/2} \leftarrow 33$s$_{1/2}$ and 32s$_{1/2} \rightarrow 31$p$_{3/2}$ transitions by monitoring the Cs$^+$-ion signal at the corresponding TOF positions as illustrated by the spectra displayed as dotted lines in Fig.\,\ref{fig:mmspecs}b) and c), respectively. The lines in these spectra have a near-Fourier-transform-limited width of 2\,MHz.

The Cs$^+$-ion TOF spectrum obtained at the position of the 32s$_{1/2}33$s$_{1/2}$ pair resonance and a delay of 1\,$\mu$s between the laser and the electric-field pulse is displayed as red trace in the upper panel of Fig.\,\ref{fig:tofs}. It consists of an intense sharp peak at early times which is attributed to Cs$^+$ ions produced before the electric-field pulse (the dynamics of the ionization process is discussed in Section\,III\,D), and a broad distribution with only weak maxima at the positions expected for the pulsed-field ionization of the atomic 32s$_{1/2}$ and 33s$_{1/2}$ states.

Compared to the TOF spectra recorded for the noninteracting Cs Rydberg gas, the state selectivity of the pulsed-field-ionization process is reduced. The reduction is attributed at least in part to adiabatic traversals of the curve crossings between different pair states, e.g. between the 32s$_{1/2}33$s$_{1/2}$ and 32p$_{3/2}32$p$_{3/2}$ states (see Fig.\,\ref{fig:foerster}) before and as the electric field is ramped up. A similar observation was made by Han and Gallagher in their study of the field-ionization dynamics in an ultracold Rb Rydberg gas (see in particular their Fig.\,4 in Ref.\,\cite{han2008}). The reduced state selectivity of the field ionization causes a millimeter-wave-radiation-independent background signal at the TOF positions corresponding to the 31p$_{3/2}$ and 33p$_{3/2}$ Rydberg states, which makes the recording of millimeter-wave spectra of the $(31$p$_{3/2}33$s$_{1/2}, 32$s$_{1/2}33$p$_{3/2})  \leftarrow 32$s$_{1/2}33$s$_{1/2}$ transitions difficult. However, comparing the Cs$^+$-ion TOF spectrum recorded with and without millimeter-wave radiation under otherwise identical conditions reveals a weak enhancement of the 33p$_{3/2}$ PFI signal and a weak depletion of the 33s$_{1/2}$ PFI signal when the millimeter-wave frequency is tuned to the 33p$_{3/2} \leftarrow 33$s$_{1/2}$ resonance (see Fig.\,\ref{fig:tofs}).

The ion-signal ratios at the TOF positions of the 33p$_{3/2}$ and 33s$_{1/2}$  states and at the positions of the 31p$_{3/2}$  and 32s$_{1/2}$  states were used to record millimeter-wave spectra of the transitions from the 32s$_{1/2}33$s$_{1/2}$ pair state to regions near the 32s$_{1/2}33$p$_{3/2}$ and 31p$_{3/2}33$s$_{1/2}$ asymptotes, respectively. These spectra are presented as solid lines in Fig.\,\ref{fig:mmspecs}. The observation of a millimeter-wave-dependent PFI signal corresponding to the  31p$_{3/2}$ and the 33p$_{3/2}$ Rydberg states represents a further experimental verification of the 32s$_{1/2}33$s$_{1/2}$ nature of the laser-excited pair resonance. Compared to the millimeter-wave spectra recorded in the noninteracting Rydberg gas, the transitions are broadened and degraded toward lower (higher) frequencies in the case of the
32s$_{1/2}33$s$_{1/2}  \rightarrow 31$p$_{3/2}33$s$_{1/2}$
(32s$_{1/2}33$p$_{3/2}  \leftarrow 32$s$_{1/2}33$s$_{1/2}$) transition. Although the large background signal and the poor signal-to-noise ratio of the millimeter-wave spectra of the pair states prevents a quantitative analysis, the spectra enable us to draw the following qualitative conclusions: i) The weak line broadenings and frequency shifts compared to the transitions recorded in the noninteracting Rydberg gas suggest that the Rydberg-atom pairs contributing to the spectra have a large interatomic separation. ii) The red (blue) shift of the transitions to the 31p$_{3/2}33$s$_{1/2}$ (32s$_{1/2}33$p$_{3/2}$) pair state reveals that the potential curves of the $n'$p$_{3/2}n$s$_{1/2}$ pair states are less strongly attractive, at long range, than is the case for the 32s$_{1/2}33$s$_{1/2}$ pair state, despite the fact that the dipole-dipole interaction is resonant, because $n$s$_{1/2}n_2$p$_j$ and $n_1$p$_{j}n$s$_{1/2}$ are degenerate when $n_1=n_2$. This aspect is discussed in more detail in Section\,IV\,D.

The observation of a prompt ionization signal further shows that an important fraction of the initial population in the 32s$_{1/2}33$s$_{1/2}$ pair state decays by ionization on the sub-microsecond timescale and therefore does not contribute to the millimeter-wave spectra. The millimeter-wave spectra can be adequately modeled (see red dashed traces in Fig.\,\ref{fig:mmspecs}) by taking into account the motion of the Rydberg-atom pairs on their potential-energy curves, as explained in Section\,IV\,D.

\begin{figure*}
\begin{center}
\includegraphics[width=0.9\textwidth]{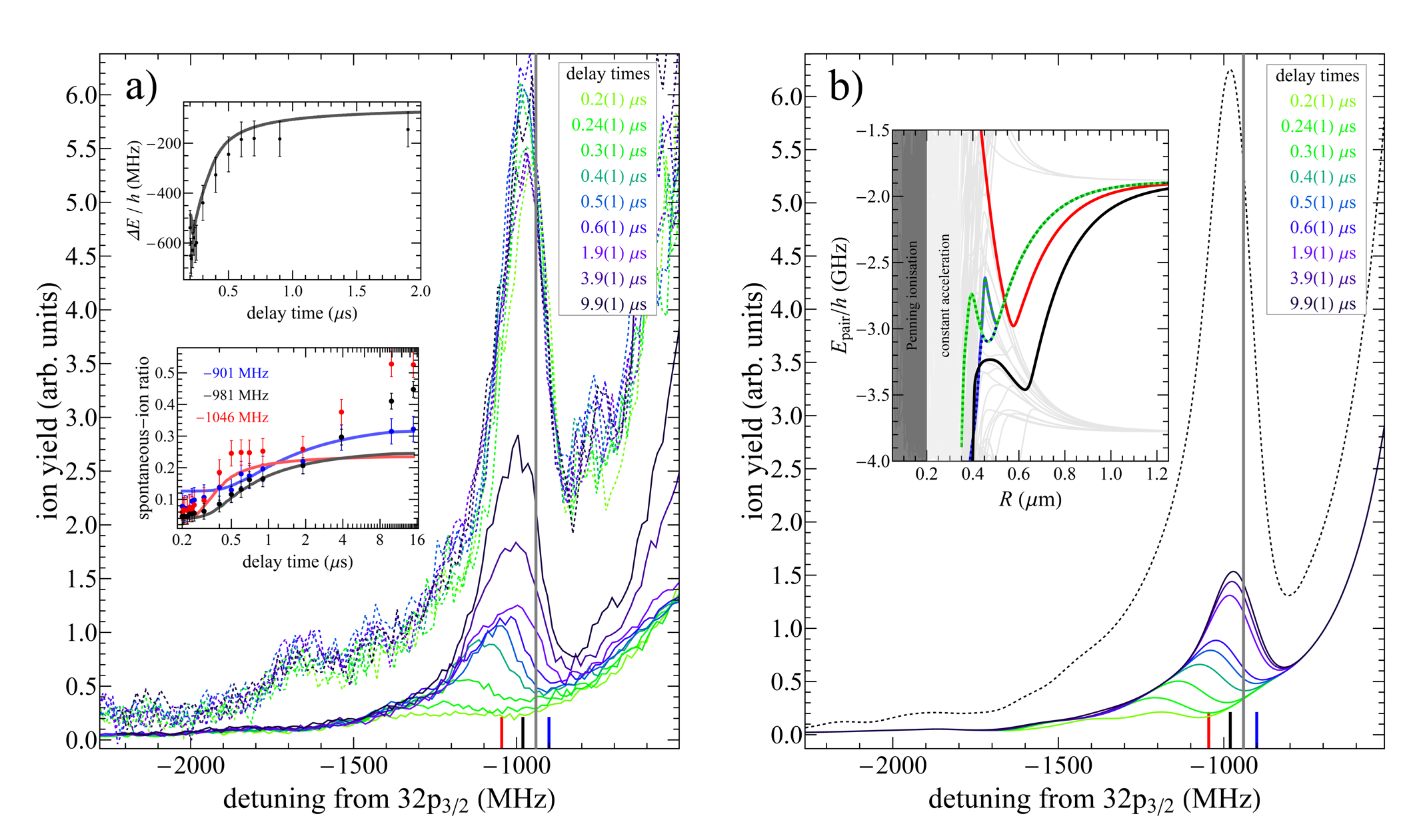}
\caption{(Color online) a) Spontaneous-ionization signal as a function of the detuning of the UV-laser frequency from the 32p$_{3/2}\leftarrow$6s$_{1/2}$ transition and of the delay between laser pulse and pulsed ion-extraction field (solid colored traces, values for the delay time given in the upper right inset). The spectral position corresponding to the asymptotic energy of the 32s$_{1/2}33$s$_{1/2}$ pair state is shown by a vertical grey bar. Simultaneously recorded spectra of the total ion signal including ions resulting from pulsed-field ionization are shown as dashed colored traces on the same scale. The interaction-induced shift (i.e., twice the detuning of the resonance peak from the asymptotic energy of the 32s$_{1/2}33$s$_{1/2}$ pair state) is given in the upper left inset and the fractions of spontaneous ion to the total-ion signals at the detunings indicated by red, black, and blue lines on the horizontal axis are displayed in the lower left inset, as functions of the delay time. Experimental values are shown as dots with error bars and simulated quantities as solid lines.  b) Spectra simulated for the same experimental parameters as in panel a), as described in Sec.\,IV\,C. The inset shows the potential-energy functions (the two $\Omega = 0$ (green, blue, and black) and the $\Omega=1$ (red) potentials associated with the 32s$_{1/2}33$s$_{1/2}$ asymptote are shown in color) of pair states below the 32p$_{3/2}32$p$_{3/2}$ asymptote as a function of the internuclear separation.}
\label{fig:delay2}
\end{center}
\end{figure*}

\subsection{Penning ionization of the 32s$_{1/2}33$s$_{1/2}$ pair state}

The spontaneous ionization of Rydberg atoms in the gas phase is the first step in the evolution of a dense Rydberg gas towards a plasma. Spontaneous ionization of a Rydberg gas was first reported by Vitrant \textit{et al.}\,\cite{vitrant1982ioni}, who pointed out that it reduces the lifetimes of Rydberg states in dense samples. Interest in the ionization process was renewed with experiments on ultracold atoms which provided detailed information on the evolution from an ultracold Rydberg gas to an ultracold plasma\,\cite{robinson2000}. This evolution can be described by an avalanche-ionization model\,\cite{pohl2003} that involves an initial seed ionization process, during which ions are produced in the ultracold gas and the free electrons escape. Once the positive space charge of the ions is sufficiently high to trap low-energy electrons, the remaining Rydberg atoms are rapidly ionized by collisions with the electrons. For the seed-ionization process, experiments pointed at the importance of a small fraction of hot Rydberg atoms and blackbody-radiation-induced photoionization~\cite{robinson2000}, static orbital overlap between close-lying Rydberg-atom pairs, sequences of quasiresonant dipole-dipole transitions~\cite{li2004,tanner2008}, and interaction-induced motion leading to Rydberg-Rydberg collisions and Penning ionization~\cite{li2005,viteau2008,reinhard2008}.

The dynamics leading to the seed ionization depends on the density of the excited Rydberg atoms, on the Rydberg state and on the details of the interaction between the Rydberg atoms. At high $n$ values ($n \approx 100$) and high Rydberg-atom densities ($>10^9$\,cm$^{-3}$), the complete ionization of the Rydberg gas occurs in less than 100\,ns, as reported in Ref.\,\cite{han2008} and confirmed in the present work (not shown). The theoretical understanding of this fast ionization process is still incomplete, but a classical simulation\,\cite{Robicheaux2014} sugggests that many-body interactions and excitation of closely-spaced pairs of Rydberg atoms play an important role under these conditions. At lower densities and $n$ values, Penning-ionization processes\,\cite{Penning1927,li2005,amthor2007,viteau2008,reinhard2008} involving two Rydberg atoms, one atom undergoing a transition to a lower state and the other being ionized, have been identified as the dominant seed-ionization mechanism.

To characterize the processes leading to the strong prompt Cs$^+$ ionization signal in our studies of millimeter-wave transitions between Rydberg-pair states (see upper panel of Fig.\,\ref{fig:tofs}), the ionization dynamics of the 32s$_{1/2}33$s$_{1/2}$ pair state was studied at a time resolution of 100\,ns limited by the temporal resolution of our Cs$^+$-ion--TOF spectra. The experiments consisted in monitoring
the yield of Cs$^+$ ions produced spontaneously during the delay time between the laser excitation pulse and the application of the pulsed electric field as a function of the UV-laser frequency. The spontaneously formed ions are accelerated from the beginning of the electric-field ramp used for the field-ionization and thus arrive first on the detector, see the integration-windows labeled ``prompt ions'' in Fig.~\ref{fig:tofs}.

The results of this investigation are presented in Fig.\,\ref{fig:delay2}a), which compares spectra of the 32s$_{1/2}33$s$_{1/2}$ pair resonance recorded for several delay times between 200\,ns and 9.9\,$\mu$s. At the shortest delay times, the peaks in the prompt-ion signal are observed $\approx2\times300$\,MHz below the 32s$_{1/2}33$s$_{1/2}$ asymptote and are very weak. The intensity of the resonance grows with increasing delay time and its position gradually shifts toward the 32s$_{1/2}33$s$_{1/2}$ asymptote, and stabilizes at $\approx2\times75$\,MHz below this asymptote at delay times beyond 1\,$\mu$s. This deviation from the asymptotic pair energy is the interaction-induced shift $\Delta E_{\rm int}$ discussed in section~\ref{sec:expExtPairstates} and shown in the inset of Fig.~\ref{fig:ndependence}. The time dependence of the resonance position is displayed in the upper left inset of Fig.\,\ref{fig:delay2}.

Because the total density of Rydberg atoms in our experiments is still very low (on the order of $1\cdot10^8$ atoms/cm$^{3}$), the experimental observations can be qualitatively understood as resulting from the purely pair-wise Penning-ionization mechanism. Pairs of Rydberg atoms are excited at well-defined initial separations given by the detuning of the excitation laser from the dissociation asymptote and are accelerated on the attractive potential-energy curves. The first pairs that ionize are those characterized by the strongest shifts, i.e., those that correspond to the shortest internuclear distances and are thus subject to the strongest attractive interactions. As the delay time increases, atom pairs subject to smaller shifts and thus weaker attractive interactions start contributing to the prompt-ion signal in the TOF spectra. These pairs have larger initial internuclear separations and require longer times before the motion on the attractive potential leads to ionizing collisions. A more quantitative analysis of this process is presented in Section\,IV\,C.

\begin{figure*}
\begin{center}
\includegraphics[width=0.9\textwidth]{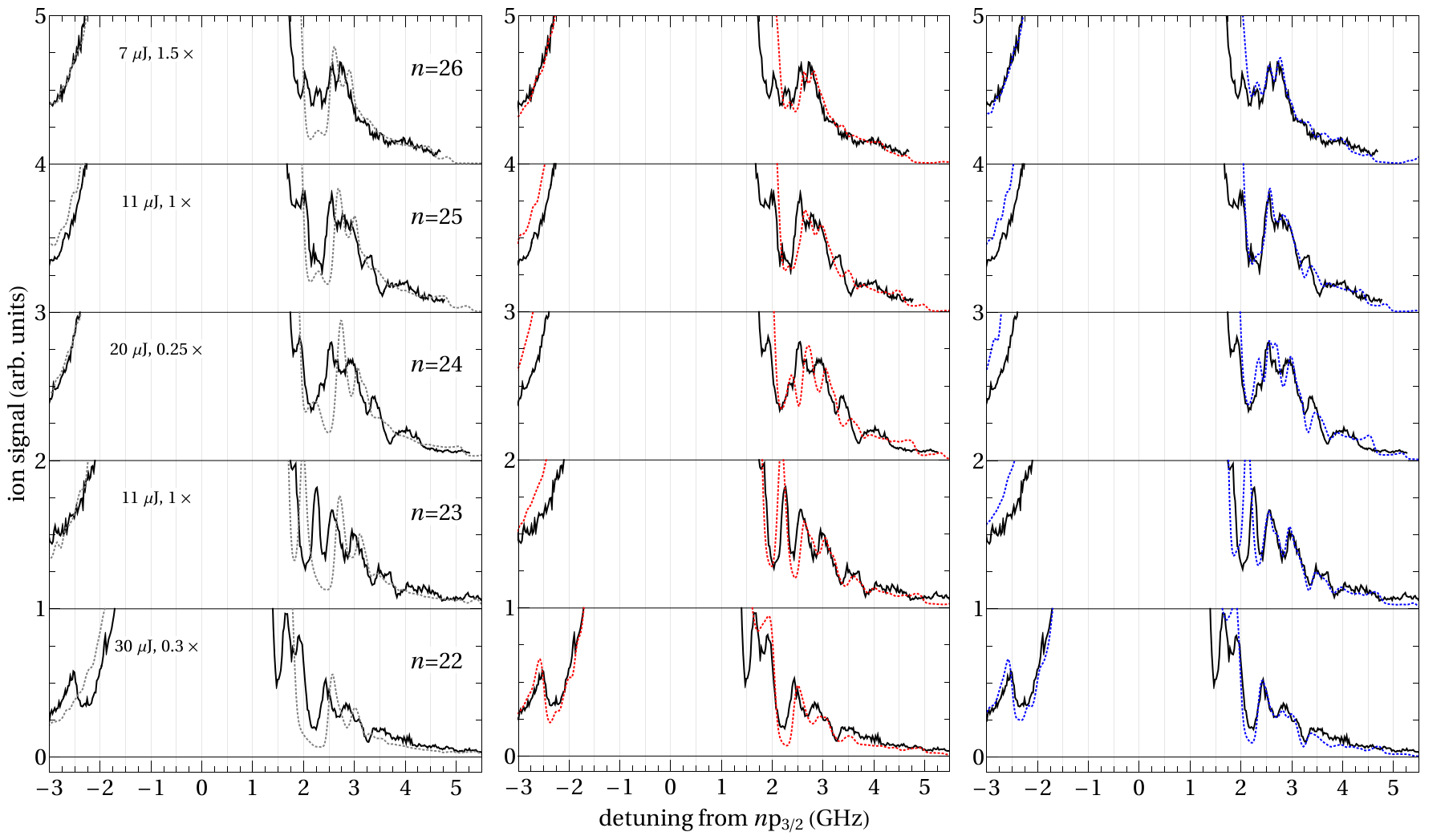}
\caption{(Color online) Spectral structures observed on the high-frequency side of the $n$p$_{3/2}n$p$_{3/2}$ ($n=22-26$) dissociation asymptotes in the UV excitation spectrum of Cs. The experimental spectra (full black lines, identical in all panel) represent the Cs$^+$-ion signal as a function of the detuning of the UV-laser frequency from the $n$p$_{3/2} \leftarrow 6$s$_{1/2}$ transition. The pulse energies used to record these spectra and the relative intensity-scaling factors are indicated below the experimental traces on the left-hand side of the left panel. The panels also display simulated spectra (dashed lines) obtained by considering all interaction terms in the multipole expansion up to dipole-dipole (left, gray), quadrupole-quadrupole (middle, red), and octupole-octupole (right, blue) interaction terms.}
\label{fig:blueside}
\end{center}
\end{figure*}

\subsection{Spectral structures originating from pair states on the high-frequency side of the \ $n$p$_{3/2}n$p$_{3/2}$ pair-dissociation asymptotes}
The interaction between the $n$p$_{3/2}n$p$_{3/2}$ and the $n$s$_{1/2}(n+1)$s$_{1/2}$ pair states schematically represented in Fig.\,\ref{fig:mo}a) leads to predominantly attractive potential-energy functions correlated to the $n$s$_{1/2}(n+1)$s$_{1/2}$ dissociation asymptotes and repulsive potential-energy functions correlated to the $n$p$_{3/2}n$p$_{3/2}$ asymptotes for the considered range of principal quantum numbers $n$. Whereas the former lead to isolated resonances on the low-frequency side of the strong atomic resonances (see, e.g., Figs.\,\ref{fig:density} and\,\ref{fig:allspectra}), the latter lead to dense and complex spectral structures on the high-frequency side of the atomic resonances, which are partially resolved at the lowest $n$ values studied ($n=22-26$). The corresponding spectra are displayed on an expanded scale in Fig.\,\ref{fig:blueside}, where they are compared to numerical simulations performed using the interaction model presented in Section\,IV\,A.

The observed spectral patterns, which are perfectly reproducible in the experiments, are more complex than those observed on the low frequency side and do not reveal any obvious scaling with the principal quantum number. They originate from the interactions of the repulsive $n$p$_{3/2}n$p$_{3/2}$ pair states with attractive $(n-4)$f$_j(n-3)$f$_j$ pair states which have their dissociation asymptotes located just above the $n$p$_{3/2}n$p$_{3/2}$ asymptote. The numerical simulations of the spectra in these regions indicate that the spectral structures do not only arise from an accumulation of spectral intensity in regions where the potential-energy functions are flat as in case of the $n$s$_{1/2}(n+1)$s$_{1/2}$ resonances, but also from a reduction of spectral intensity centered around avoided crossings between states correlated to the $n$p$_{3/2}n$p$_{3/2}$ and $(n-4)$f$_j(n-3)$f$_j$ dissociation asymptotes. This aspect will be discussed in more detail in Section\,IV\,B.

\section{Potential model for interacting Rydberg-atom pairs and interpretation of experimental results}
\subsection{Model of long-range potential-energy curves}
The potential model we have developed to analyze the results presented in Section\,III was briefly introduced in Ref.\,\cite{deiglmayr2014} and is presented in detail here. We describe the long-range interaction between the two neutral atoms A and B in the frame of the nonrotating molecule AB using a multipole expansion of the form\,\cite{flannery2005}
\begin{align}\label{eq:vint}
V_\textrm{inter} & =\nonumber \\  \sum_{\mathclap{L_{\AR/\BR}=1}\hspace{0.5cm}}^\infty &\sum_{\hspace{0.5cm}\mathclap{\omega=-L_<}}^{+L_<}\frac{(-1)^{L_\BR}f_{L_\AR,L_\BR,\omega}}{R^{L_\AR+L_\BR+1}} Q_{L_\AR,\omega}(\vec{r}_\AR)Q_{L_\BR,-\omega}(\vec{r}_\BR)  \\
Q_{L,\omega}(\vec{r}) &= \sqrt{\frac{4 \pi}{2 L +1}} r^L Y_{L,\omega}(\hat{r}) \label{eq:multipoles} \\
f_{L_\AR,L_\BR,\omega} & =  \frac{(L_\AR+L_\BR)!}{\sqrt{(L_\AR+\omega)!(L_\AR-\omega)!(L_\BR+\omega)!(L_\BR-\omega)!}},
\end{align}
where $\vec{r}_i$ is the position of the Rydberg electron bound to atom $i$ with respect to its core, $R$ the interatomic separation, $Q_{L_i\omega}(\vec{r}_i)$ the atomic multipole of $L_i$-th order ($L_<$ being the smaller of two $L_i$ values), and $\omega$ is the quantum number associated with the projection of the angular momentum $L_<$ onto the molecular axis. This expansion is valid in the absence of retardation effects, which are negligible in the systems discussed here, and for interatomic separations exceeding the LeRoy-radius $R_\textrm{LR}=2\left(\braket{r_\AR^2}^{1/2}+\braket{r_\BR^2}^{1/2}\right)$~\cite{leroy1973}, a criterion specifying the internuclear distance beyond which orbital overlap between the two atoms becomes negligible. The atomic multipoles $Q_{L_i\omega}(\vec{r}_i)$ are expressed in terms of spherical harmonics $Y_{L,\omega}$ (see Eq.\,(\ref{eq:multipoles})).
To account for the experimental observations, we found it sufficient to restrict the multipole expansion to contributions up to the octupole-octupole ($L_{\rm A}=3, L_{\rm B}=3$) term. The Hamiltonian
\begin{equation}
H =H^\AR_0+H^\BR_0+V'_\textrm{inter}
\label{eq:htot}
\end{equation}
consists of the Hamiltonians $H^{\rm A,B}_0$ of the two isolated atoms taking into account the quantum defects of all states with $l \leq 4$ (from Refs.\,\cite{goy1982,weber1987}) and the truncated multipole expansion term $V'_{\rm inter}$.

To obtain the adiabatic potential-energy curves of the interacting Rydberg-atom pairs at distances larger than the LeRoy radius, $H$ is determined in matrix form using the product basis $\ket{n_\AR l_\AR j_\AR \omega_\AR , n_\BR l_\BR j_\BR \omega_\BR}\equiv\ket{ \gamma_\AR j_\AR \omega_\AR, \gamma_\BR j_\BR \omega_\BR}$, following the approach described in Refs.~\cite{stanojevic2008,stanojevic2006,samboy2011}. Here, $\omega_{\rm A,B}$ represent the projections of the Rydberg-electron angular momentum onto the internuclear axis, with $\omega_{\rm A}+\omega_{\rm B}=\Omega$. We neglect the hyperfine structure of the Rydberg states because the largest hyperfine splittings for the low-$n$ s$_{1/2}$ states are still more than one order of magnitude smaller than our laser bandwidth\,\cite{sassmannshausen2013}.
The angular and radial parts of the matrix elements of $H$ can be separated and the angular matrix elements are calculated analytically, whereas the radial matrix elements are obtained by numerical integration\,\cite{zimmerman1979} using experimental quantum defects~\cite{goy1982,weber1987} and an \textit{ab-initio} model potential~\cite{marinescu1994}. The details of the calculation of the matrix elements are presented in the appendix.

\begin{figure*}
\begin{center}
\includegraphics[width=0.99\textwidth]{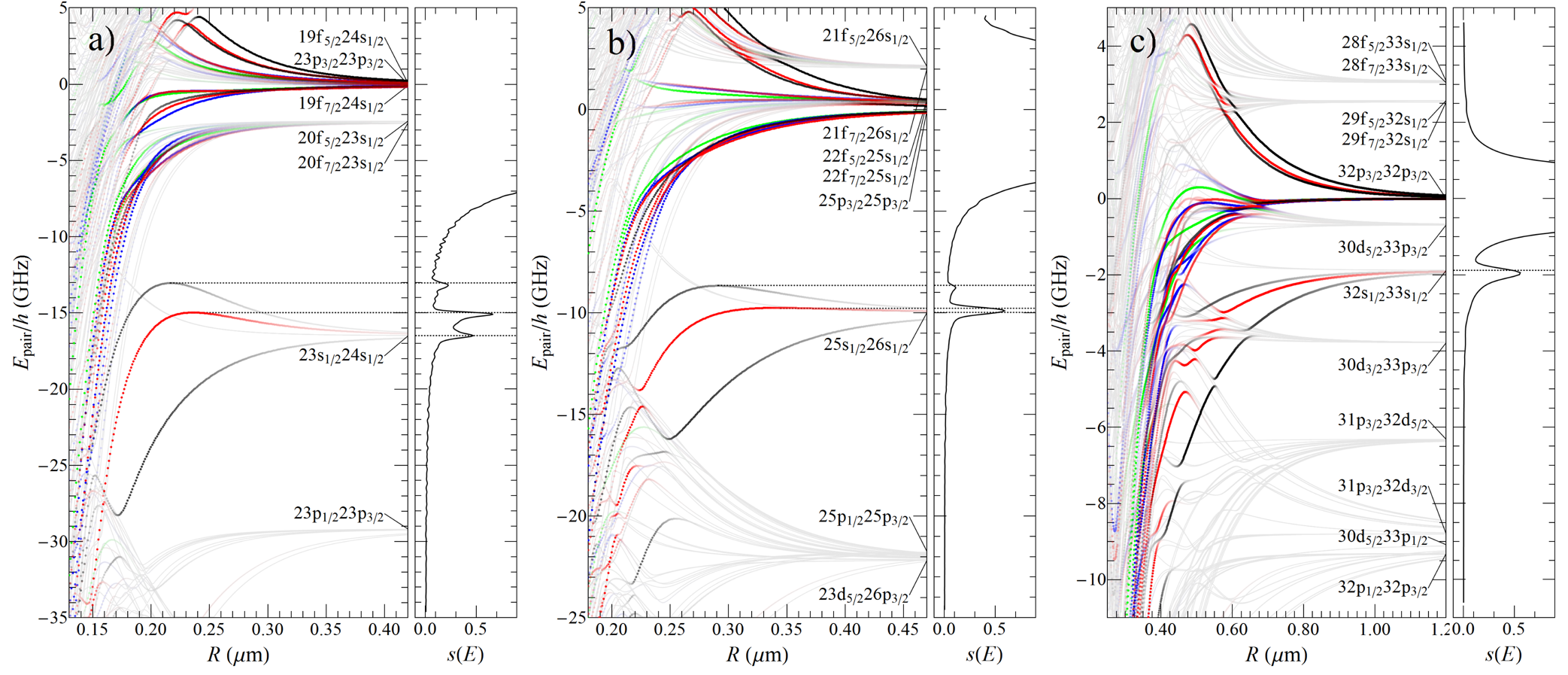}
\caption{(Color online) Calculated potential-energy curves for long-range Rydberg-atom pairs in the vicinity of $n$p$_{3/2}n$p$_{3/2}$ asymptotes at a) $n=23$, b) $n=25$, and c) $n=32$ and for $\Omega=0,1,2,3$ (black, red, blue, and green curves, respectively) as obtained from our long-range potential model (details in the text). The intensity of the color reflects the value of $\overline{p_\Psi}(R)$, gray color denoting 0\% overlap with the atomic  $n$p$_{3/2}n$p$_{3/2}$ state, full color denoting 1\% or higher overlap. The simulated spectra plotted in arb. units on the right-hand side of each of the three panels were calculated using Equation\,(\ref{eq:lineprofile}).}
\label{fig:pecs}
\end{center}
\end{figure*}

\begin{figure}
\begin{center}
\includegraphics[width=0.95\linewidth]{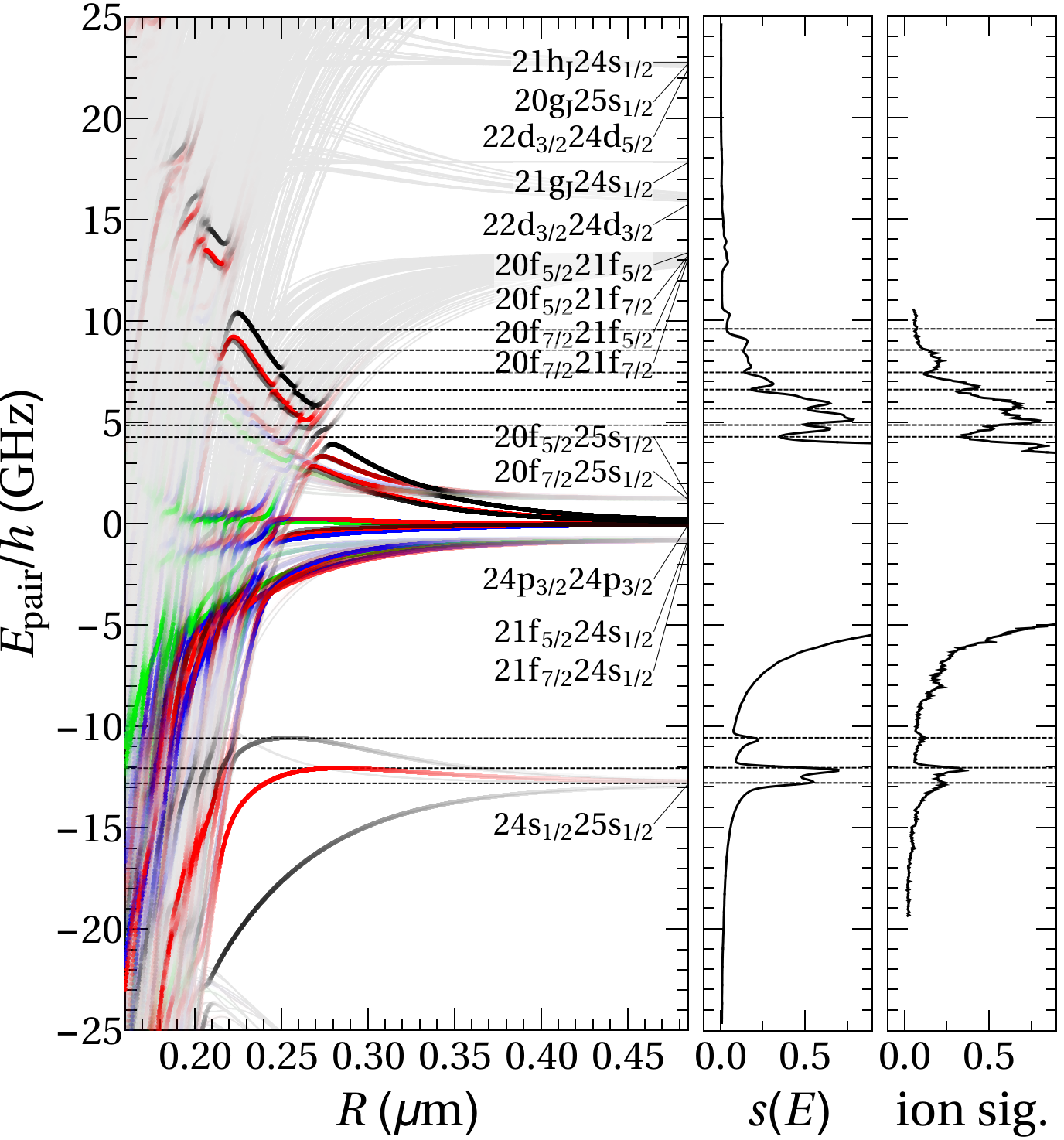}
\caption{(Color online) Calculated potential-energy functions near the 24p$_{3/2}24$p$_{3/2}$ pair-dissociation asymptote (left panel), simulated spectrum $s(E)$ (middle, arb. units), and experimentally measured Cs$^+$-ion signal as a function of the two-photon detuning $2\Delta \nu$ (right panel, arb. units). The color code is the same as in Fig.~\ref{fig:pecs}. The horizontal dashed lines indicate the positions of minima and maxima in the simulated spectrum for comparison with the experimental spectrum and the calculated potential-energy functions.}
\label{fig:pec}
\end{center}
\end{figure}

The quantum number $\Omega$ associated with the projection of the total electronic angular momentum onto the molecular axis is a good quantum number.
As pointed out in Ref.\,\cite{deiglmayr2014}, although coupling terms with $(-1)^{L_{\rm A}+L{\rm B}}=-1$, such as the dipole-quadrupole coupling term, do not conserve the parity of the electronic wave function, they nevertheless lead to observable interactions because the coupling of quasi-degenerate rotational levels ensures the conservation of the total (rovibronic) parity. In Ref.\,\cite{deiglmayr2014}, we erroneously also stated that such terms can be accompanied by a mixing of states of {\textit{gerade} (g)} and {\textit{ungerade} (u)} electronic symmetry in homonuclear dimers. The {\it g-u} mixing in Cs is actually induced by hyperfine interactions, which are neglected in our model.
The eigenvalues of the Hamiltonian $H$ are determined in an unsymmetrized basis independently for the different values of $\Omega=(0,1,2,3)$ that are optically accessible, and for different values of the internuclear separation $R$. The molecular eigenfunctions $\ket{\Psi(R)}=\sum c_{\gamma_\AR j_\AR  \omega_\AR, \gamma_\BR j_\BR \omega_\BR }(R)\ket{\gamma_\AR  j_\AR \omega_\AR, \gamma_\BR j_\BR \omega_\BR }$ and the molecular eigenenergies $E_\Psi(R)$ are determined as a function of the interatomic separation $R$. Exemplary potential-energy curves describing the long-range interaction between two Cs Rydberg atoms are displayed in Figs.\,\ref{fig:pecs} and\,\ref{fig:pec}.

The density of Rydberg-atom pair states is very high because of the large number of pair dissociation asymptotes and because several potential-energy functions correlate to each asymptote. Only a few of the multitude of molecular pair states are accessible optically  in our experiments. The excitation probability $\overline{p_\Psi}(R)$ of a molecular eigenstate $\ket{\Psi}$ at a pair separation $R$  is proportional to the square of the mixing coefficients of the $n$p$_{3/2}n$p$_{3/2}$ basis states. The  optically accessible states $\ket{\phi}$ (for $n$p$_{3/2}n$p$_{3/2}$ we take $n$,$l=1$,$j_{\rm A}=j_{\rm B}=3/2$) are conveniently expressed in the laboratory-fixed frame as
\begin{align}
\ket{\phi}=\sum_{m_\AR, m_\BR} a_{j_\AR m_\AR, j_\BR m_\BR}\ket{j_\AR m_\AR, j_\BR m_\BR}.
\label{eq:atomicstate}
\end{align}
To account for our experimental conditions, we assume an isotropic distribution of 6s$_{1/2}$ ground-state Cs atoms ($m_s=\pm 1/2$). $\sigma^+$ polarized transitions thus lead to the excitation of four different states $\phi$ corresponding to $j_{\AR}=j_{\BR}=3/2$ and $m_\AR=1/2,3/2$ and $m_\BR=1/2,3/2$ with equal weights. The intensity is obtained as the squared overlap $p_\Psi(\theta,R)=\left|\braket{\Psi(R)|\phi}\right|^2$ between the optically accessible function $\phi$ and the function $\Psi(R)$ expressed in the molecule-fixed frame with the internuclear axis forming an angle\,$\theta$ with the laboratory-fixed $z$ axis. To calculate $p_\Psi(\theta,R)$, we transform the atomic functions $j_{\rm A,B}m_{\rm A,B}$ into the molecule-fixed frame using the Wigner d rotation matrices and obtain
\begin{align}
p_\Psi(\theta,R)&=\left|\braket{\Psi(R)|\phi}\right|^2= \nonumber \\
&\Big|\sum_{\mathclap{\substack{\gamma_\AR j_\AR  \omega_\AR, \gamma_\BR j_\BR \omega_\BR  \\ j_\AR m_\AR, j_\BR m_\BR}}}\braket{\Psi(R)|\gamma_\AR j_\AR  \omega_\AR, \gamma_\BR j_\BR \omega_\BR } d_{\omega_{\AR} m_{\AR}}^{j_{\AR}}(\theta) \times \nonumber
\\ & \hspace{1.5cm} d_{\omega_{\BR} m_{\BR}}^{j_{\BR}}(\theta) \braket{j_\AR m_\AR, j_\BR m_\BR|\phi}\Big|^2 \hspace{0.2cm}.
\label{eq:rot}
\end{align}
We average $p_\Psi(\theta,R)$ over all orientations $\theta$ of the molecular axis relative to the $z$ axis and obtain $\overline{p_\Psi}(R)=\int p_\Psi(\theta,R)2 \pi {\rm sin}(\theta) d \theta$, which is directly proportional to the spectral intensity resulting from excitation of a molecular pair state $\ket{\Psi}$ at the interatomic separation $R$. The corresponding spectral line profile $s_\Psi(E')$ is then determined as
\begin{multline}
\label{eq:lineprofile1}
    s_\Psi(E') = \\
     4 \pi \int_0^\infty \delta(E'-E_{\Psi}(R)) \frac{4\, \omega_\textrm{atom}^4}{(E_{\rm pp}-E_{\Psi}(R))^2} \overline{p_\Psi}(R) R^2 \textrm{d}R,
\end{multline}
where $\omega_\textrm{atom}$ is the Rabi frequency of the atomic $n$p$_{3/2}\leftarrow6$s$_{1/2}$ transition, $E_{\rm pp}$ the asymptotic energy of the $n$p$_{3/2}n$p$_{3/2}$ state, and the term $4\, \omega_\textrm{atom}^4 /(E_{\rm pp}-E_{\Psi}(R))^2$ accounts for the two-photon excitation of the pair-state with an off-resonant intermediate $n$p$_{3/2}6$s$_{1/2}$ (or 6s$_{1/2}n$p$_{3/2}$) state at a detuning of $\frac{1}{2}\left(E_{\rm pp}-E_{\Psi}(R)\right)$  \cite{stanojevic2008}. The delta function ensures resonance with the molecular state with the energy $E_{\Psi}(R)$ and intensity $\overline{p_\Psi}(R)$. Assuming a gas of homogeneous density, the number of atom pairs at a separation $R$ is proportional to $4 \pi R^2$. For comparison with experimental spectra, the line profile $s_\Psi(E')$ is convoluted with the laser line profile $G(E)$ and summed over all molecular states $\ket{\Psi}$ to obtain the full spectrum $s(E)$ as
\begin{align}
\label{eq:lineprofile}
    s(E) &=\sum_\Psi \int_{-\infty}^\infty s_\Psi(E') \cdot G(E-E') \textrm{d}E' \nonumber \\
&= 16 \pi \sum_\Psi \int_{-\infty}^\infty \int_0^\infty \delta(E'-E_{\Psi}(R)) \frac{\omega_\textrm{atom}^4}{(E_{\rm pp}-E_{\Psi}(R))^2} \times \nonumber \\
& \hspace{1.5cm} \overline{p_\Psi}(R) R^2 \cdot G(E-E') \textrm{d}R\textrm{d}E'   \nonumber \\
&=16 \pi \sum_\Psi \int_0^\infty  \frac{\omega_\textrm{atom}^4}{(E_{\rm pp}-E_{\Psi}(R))^2} G(E-E_\Psi(R)) \times \nonumber \\
& \hspace{1.5cm} \overline{p_\Psi}(R) R^2 \textrm{d}R.
\end{align}
The laser line profile $G(E)$ is assumed to be a Gaussian function with the experimentally determined full width at half maximum of 140\,MHz. The atomic Rabi frequencies $\omega_{\rm atom}$ of the single-photon $n$p$_{3/2}\leftarrow6$s$_{1/2}$ transitions act only as distance- and energy-independent overall scaling factors and are not taken explicitly into account when fitting the global amplitude of the simulated spectra $s(E)$ to the experimental data (see Figs.\,\ref{fig:allspectra} and\,\ref{fig:blueside}).

The intensity of the color used to draw the potential-energy functions in Fig.\,\ref{fig:pecs} represents the value of the function $\overline{p_\Psi}(R)$, gray corresponding to a negligible contribution of the optically accessible $n$p$_{3/2}n$p$_{3/2}$ component and full color to a $n$p$_{3/2}n$p$_{3/2}$ contribution of at least 1\%.

At large distances, the only nonvanishing contributions are located in the immediate vicinity of the $n$p$_{3/2}n$p$_{3/2}$ asymptotes. At shorter distances, many more potential curves contribute to the spectral intensity. However, most potential curves are strongly attractive and the large gradients spread the overall intensity over broad spectral ranges, which renders the observation of pair states impossible. From the general appearance of the curves depicted in Fig.\,\ref{fig:pecs}, one expects sharp spectral structures to arise from regions where the potential curves that possess some $n$p$_{3/2}n$p$_{3/2}$ character are flat. Such regions of enhanced intensity are encountered where the curves have a local maximum (a clear example of such a situation is given by the red and black curves correlated to the 23s$_{1/2}24$sp$_{1/2}$ asymptote in the left panel of Fig.\,\ref{fig:pecs}) or closely approach their asymptotic value (all three curves correlated to $n$s$_{1/2}(n+1)$s$_{1/2}$ at large $R$ values in Fig.\,\ref{fig:pecs}).

Sharp spectral structures are also encountered in regions where avoided crossings reduce the spectral density, leading to intensity minima. Characteristic examples of this situation are the minima observed in the high-frequency wings of the $n$p$_{3/2}n$p$_{3/2}$ resonances for $n=22-26$ in Fig.\,\ref{fig:blueside}.

\subsection{Interpretation of the UV-laser spectra}
The procedure we followed to interpret the experimental UV-laser spectra of Rydberg-atom pairs such as those presented in Figs.\,\ref{fig:density}, \,\ref{fig:allspectra} and \ref{fig:blueside} was to evaluate the spectrum $s(E)$ using Eq.\,(\ref{eq:lineprofile}) retaining successively higher terms of the long-range interaction series (Eq.\,(\ref{eq:multipoles})), which, in turn, necessitated the extension of the basis set $\ket{n_\AR l_\AR  j_\AR \omega_\AR, n_\BR l_\BR j_\BR \omega_\BR}$ to include pair states with $l_\AR$ and $l_\BR$ values up to 6. Convergence of the eigenvalue calculations was checked by increasing the range of principal quantum numbers $n_\AR$ and $n_\BR$. Because the atomic radial transition matrix elements rapidly decrease with the value of $\Delta n$, only states differing in effective principal quantum number $n^*$ by no more than 1 were considered in the calculations. It further turned out to be sufficient to only consider the pair states with asymptotes located within an energy window of $0.64 \cdot 4 R_{\textrm{Cs}}/(n^*)^3$ around the relevant $n$p$_{3/2}n$p$_{3/2}$ asymptote. States located outside this range only approach the $n$p$_{3/2}n$p$_{3/2}$ potential curves at short range, where the gradients of the potential-energy curves are large and spectral structures are broad and weak.

As explained in Section\,IV\,A, the model does not include adjustable parameters, but relies on the known quantum defects of the Cs-atom Rydberg states. The structures of the spectra in the regions located below the $n$p$_{3/2}n$p$_{3/2}$ dissociation asymptote (see Fig.\,\ref{fig:allspectra}) are primarily determined by the dipole-dipole interaction term which couples the $n$p$_{3/2}n$p$_{3/2}$ and $n$p$_{1/2}n$p$_{3/2}$ pair states with the $n$s$_{1/2}(n+1)$s$_{1/2}$ pair states. The potential-energy functions that give the $n$s$_{1/2}(n+1)$s$_{1/2}$ resonances their characteristic interaction-induced shifts (see inset of Fig.\,\ref{fig:ndependence}) and fine structures (see Fig.\,\ref{fig:allspectra}) are depicted in Fig.\,\ref{fig:pecs} for $n=23$, 25 and 32. The calculated spectra $s(E)$ are depicted on the right-hand side of each panel, and dashed horizontal lines highlight the origins of the sharp spectral structures.

Three states correlating to the $n$s$_{1/2}(n+1)$s$_{1/2}$ asymptotes possess a significant admixture of the optically accessible $n$p$_{3/2}n$p$_{3/2}$ pair state, two of which are $\Omega=0$ (black lines) and the third (red line) is a $\Omega=1$ state.
At the highest $n$ values studied experimentally ($27 < n < 37$), for which the results obtained at $n=32$ are representative (see Fig.\,\ref{fig:pecs}), all three states are attractive. In the range of internuclear distances where the $n$p$_{3/2}n$p$_{3/2}$ character is significant, the dipole-dipole interaction with the $n$p$_{3/2}n$p$_{3/2}$ state is the dominant interaction for all three states. The interaction with the $n$p$_{1/2}n$p$_{3/2}$ pair state, which is located more than 10\,GHz below the $n$p$_{3/2}n$p$_{3/2}$ asymptote at $n=32$, does not significantly affect the $n$s$_{1/2}(n+1)$s$_{1/2}$ potential curves in this range of $n$ values. The calculations predict a red detuning of the pair resonance with respect to the $n$s$_{1/2}(n+1)$s$_{1/2}$ asymptote and a red-degraded line shape, in excellent agreement with the experimental observations (see Fig.\,\ref{fig:allspectra}).

As $n$ decreases below $n=28$, the position of the $n$s$_{1/2}(n+1)$s$_{1/2}$ asymptotes rapidly shifts away from the $n$p$_{3/2}n$p$_{3/2}$ asymptote and toward the  $n$p$_{1/2}n$p$_{3/2}$ asymptote. Consequently, the effect of the dipole-dipole interaction with the $n$p$_{1/2}n$p$_{3/2}$ pair state becomes important: The $\Omega=1$ and one of the two $\Omega=0$ potential functions become repulsive at intermediate distances beyond about 0.3\,$\mu$m at $n=25$ and beyond about 0.20\,$\mu$m at $n=22$. At shorter range, the interaction with the $n$p$_{3/2}n$p$_{3/2}$ pair state is always dominant and makes the $n$s$_{1/2}(n+1)$s$_{1/2}$ potentials attractive, resulting in local maxima in these two curves. These two maxima give rise to two peaks in the calculated spectra located on the high-frequency side of the $n$s$_{1/2}(n+1)$s$_{1/2}$ asymptotes. The third peak, which corresponds to the position of the asymptote, is well resolved below $n<25$. The calculated spectra $s(E)$ accurately reproduce the experimental spectra at all $n$ values investigated (see Fig.\,\ref{fig:allspectra}). The interaction-induced shifts of the pair resonances extracted from simulated and experimental spectra agree within the experimental uncertainty, as shown in the inset of Fig.~\ref{fig:ndependence}. Even the positive interaction-induced shift of the $n=25$ pair resonance with respect to the 25s$_{1/2}26$s$_{1/2}$ asymptote (see Fig.\,\ref{fig:ndependence}) is accounted for by the calculations, which indicate that it results from the overlap of signals originating from the long-range part of the potentials and from the flat potential region near the maximum of the $\Omega=1$ curve (red trace in Fig.\,\ref{fig:pecs}b)).

The potential-energy functions of the $\Omega =0$ (black), 1 (red), 2 (blue), and 3 (green) states required to interpret the spectral structures located on the high-frequency side of the 24p$_{3/2}24$p$_{3/2}$ asymptotes are depicted in Fig.\,\ref{fig:pec}. Many pair states correlating to the 20f$_{5/2,7/2}25$s$_{1/2}$, 20f$_{5/2,7/2}21$f$_{5/2,7/2}$, 22d$_{3/2}24$d$_{3/2,5/2}$, 21g$_{j}24$s$_{1/2}$, 20g$_{j}25$s$_{1/2}$ and 21h$_{j}24$s$_{1/2}$ asymptotes are located in this region, but they only possess 24p$_{3/2}24$p$_{3/2}$ character near avoided crossings with the $n$p$_{3/2}n$p$_{3/2}$ ($\Omega=0-3$) states. In the absence of these crossings, the repulsive curves associated with the 24p$_{3/2}24$p$_{3/2}$ asymptote would lead to a broad and structureless high-frequency wing of the atomic transition. At the avoided crossings no curve possesses significant 24p$_{3/2}24$p$_{3/2}$ character. Consequently, the spectral excitation function $s(E)$ has minima that can be directly linked to the avoided crossings, as indicated by the horizontal dashed lines in Fig.\,\ref{fig:pec}.

The simulations of the spectra presented in Fig.\,\ref{fig:blueside} indicate that a potential model restricted to the dipole-dipole interaction fails to reproduce the details of the spectral structures (left panel) on the high-frequency side of the atomic transitions and that the successive inclusion of interaction terms up to the octupole-octupole term gradually improves the agreement between measured and calculated spectra.

The fact that the calculations almost perfectly reproduce the experimental spectra (see Figs.\,\ref{fig:allspectra} and \ref{fig:blueside}) is remarkable and clearly reveals that (i) the long-range adiabatic potential model is adequate both with respect to the multipole expansion and the basis set, and (ii) pairwise interactions are sufficient to fully describe the behavior of the Rydberg gas at the time scale probed by our laser pulses ($\approx 5$\,ns) under the experimental conditions used to record the spectra.

It may appear surprising at first sight that adiabatic potential-energy functions can be used to accurately describe the spectra of Rydberg-atom pairs. Indeed, the high density of electronic states implies slow electronic dynamics, which in general tends to be accompanied by a breakdown of the Born-Oppenheimer and adiabatic approximations. The reason why the potential-energy curves provide an accurate description of the spectra is that the electronic dynamics, though slow, is still several orders of magnitude faster than the nuclear dynamics. For any given value of $\Omega$, the state density of electronic Rydberg-atom pairs is approximately $1-5$\,GHz$^{-1}$ at $n=24$ which implies that the timescale of the electronic motion is about 1\,ns. At the translational temperature of 40\,$\mu$K of the Cs vapor, the mean velocity of the Cs atoms is about 10\,cm/s, so that they can be considered stationary on the timescale of the electronic motion, and also during the UV-laser pulse. At internuclear distances beyond $R=200$\,nm, the gradients of the potential curves are less than 500\,GHz/$\mu$m so that the relative velocity acquired by the atoms in 1\,ns is less than 2\,m/s which is also negligible. Over longer timescales, the acceleration of the atoms in the attractive potentials inevitably leads to nonadiabatic effects and eventually to Penning ionization.

\subsection{Simulation of Penning ionization}
To analyze the Penning-ionization experiments described in Section\,III\,D (see in particular Fig.\,\ref{fig:delay2}a), we use a simple model which relies on the spectral excitation function $s(E)$ and treats the relative motion of the atoms on the two $\Omega=0$ and the $\Omega=1$ potential curves associated with the 32s$_{1/2}33$s$_{1/2}$ asymptote classically. All other pair states are excited at short range (i.e., at $R$ values below 0.4\,$\mu$m) and only contribute to the spectra weakly on the low-frequency side of the resonance. The potential curves of the relevant molecular states are drawn in color in Fig.\,\ref{fig:delay2}b). Below 0.7\,$\mu$m, the potential-energy curves reveal several avoided crossings. Following photoexcitation, the two interacting Rydberg atoms are accelerated towards each other until Penning ionization takes place, which is assumed, for simplicity, to happen at an interatomic distance of 200\,nm. The probabilities of adiabatic traversals of the avoided crossings are estimated with the Landau-Zener formula
\begin{equation}
P_{\rm diab}={\rm exp}\left\{-2\pi\frac{|V_{12}|^2}{\hbar({\rm d}E/{\rm d}t)}\right\}; P_{\rm adiab}=1-P_{\rm diab},
\label{eq:landauzener}
\end{equation}
where $V_{12}$ represents the minimal distance between the potential curves and the slew rate ${\rm d}E/{\rm d}t$ is determined from the relative motion of the two atoms on the potential curves. The first avoided crossing of the $\Omega=1$ state (see red curve in Fig.\,\ref{fig:delay2}b)) is traversed adiabatically ($P_{\rm adiab}>99\%$). The adiabatic potential has a minimum at $R\approx0.58\,\mu$m and the state therefore does not rapidly reach the region where Penning ionization takes place. The motion on the two $\Omega=0$ potential curves (black and blue/green lines in Fig.\,\ref{fig:delay2}b)) is such that the regions of short internuclear distances where Penning ionization takes place is reached rapidly despite several avoided crossings. The simulation of the ionization dynamics of the $\Omega=0$ states adequately reproduces both the rise of the spontaneous-ionization signal and the gradual shift of the peak position (see insets of Fig.\,\ref{fig:delay2}a)) observed experimentally for delay times up to $\approx 2$\,$\mu$s. For longer delay times a slower rise of the spontaneous-ionization signal up to $\approx 10$\,$\mu$s is observed, which is not explained by our model (see lower inset of Fig.\,\ref{fig:delay2}a)). We attribute this additional ion signal to the ionization of pairs of atoms in the $\Omega=1$ state. A possible mechanism could be a weak mixing of $\Omega=0$ and $\Omega=1$ states (\textit{e.g.} by a residual electric field), which would lead to a small avoided crossing of the two symmetries (see red and green/blue traces in the inset of Fig.~8b) at $R\sim0.5$~$\mu$m) and open a path for ionization of pairs in the metastable $\Omega=1$ state. The probability for an adiabatic crossing is inversely proportional to the velocity of the two colliding atoms at the crossing (see Eq.~\eqref{eq:landauzener}). For pairs excited at lower frequencies, this velocity decreases and the ionization probability thus increases. This is consistent with the observed increase of the deviation between experimental and simulated ionization signals at longer delays when the excitation laser is tuned to lower frequencies (see lower inset in Fig.~8a)). For long delay times, the experimental spontaneous-ion signal approaches approximately 50\% of the initial total-ion signal, which indicates that every pair of atoms eventually undergoes a Penning ionization process.

The model discussed in this section predicts the correct qualitative behavior of fast and slow spontaneous ionization processes, but does not satisfactorily explain the observation of an almost constant total-ion signal for all experimental delay times (see dashed-colored traces in Fig.~\ref{fig:delay2}a)). Indeed, we predict that the Rydberg atom undergoing a transition to a lower Rydberg state in the Penning ionization process must end in a level with $n^*<21$, whereas the amplitude of the pulsed electric field used for field ionization (1250\,V/cm) should only ionize Rydberg states with $n^*>22$.

\subsection{Simulation of millimeter-wave spectra between Rydberg-atom--pair states}
\begin{figure}
\begin{center}
\includegraphics[width=0.99\linewidth]{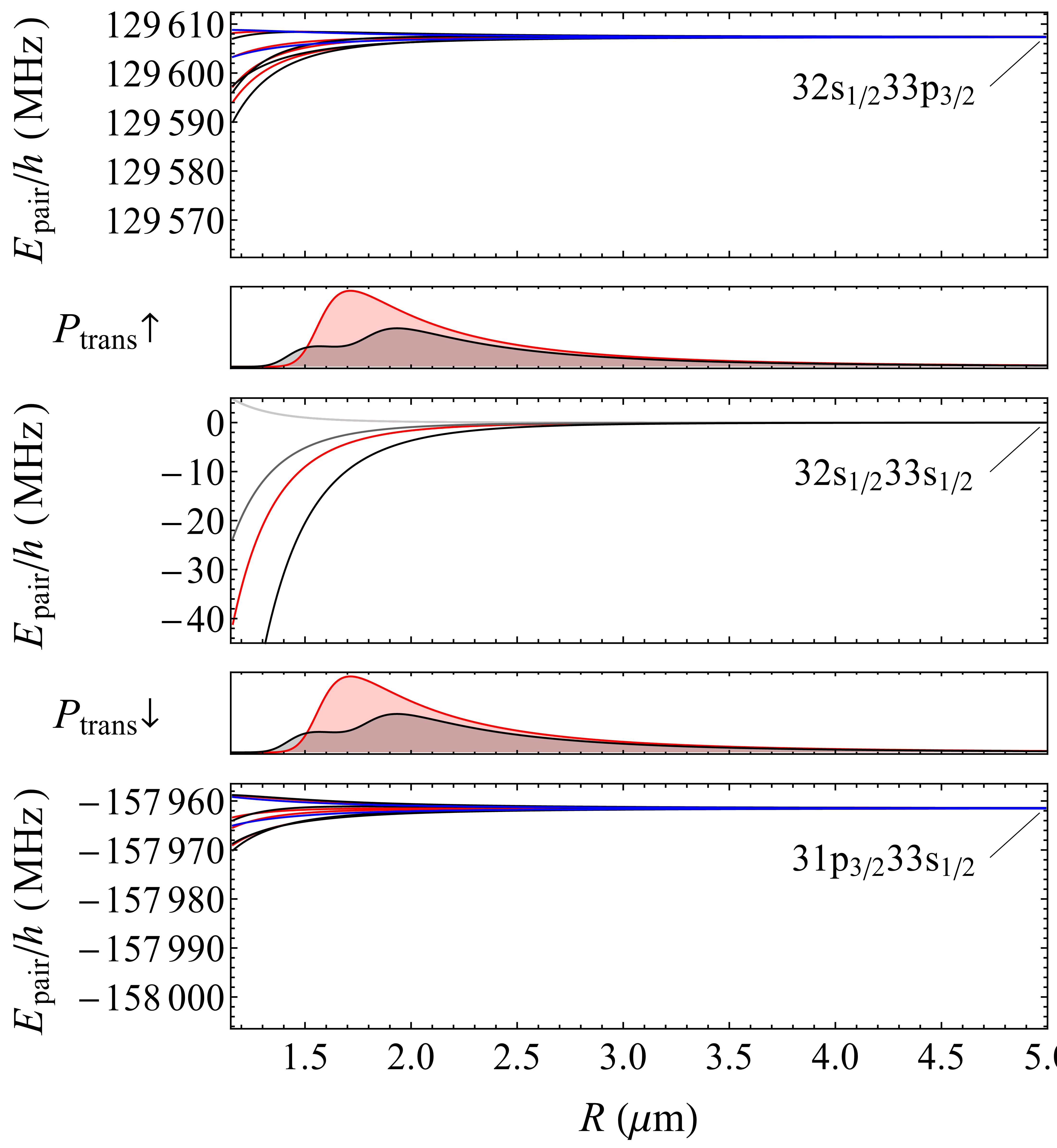}
\caption{(Color online) Potential-energy curves of the states correlated to the 32s$_{1/2}$33s$_{1/2}$  (middle panel), the 32s$_{1/2}$33p$_{3/2}$ (top panel) and the 31p$_{3/2}$33s$_{1/2}$ (bottom panel) asymptotes used for the simulation of the millimeter-wave spectra presented as red dashed lines in Figs.\,\ref{fig:mmspecs}b) and c). Black (grey), red and blue designate states with $\Omega = 0,1$ and 2, respectively. The intermediate panels present the transition probabilities calculated for the absorption ($P_{\rm trans}\uparrow$) and stimulated emission ($P_{\rm trans}\downarrow$) out of the $\Omega = 0$ (black) and $\Omega = 1$ (red) states correlated to the 32s$_{1/2}$33s$_{1/2}$ asymptote. All energies are given relative to the energy of the 32s$_{1/2}$33s$_{1/2}$ asymptote. See text for details.
}
\label{mmPEC}
\end{center}
\end{figure}

The potential model and the spectral distribution function $s(E)$ presented in Section IV A were also used to calculate the millimeter-wave spectra of the   32s$_{1/2}$33p$_{3/2} \leftarrow 32$s$_{1/2}$33s$_{1/2}$ and 32s$_{1/2}$33s$_{1/2} \rightarrow 31$p$_{3/2}$33s$_{1/2}$ transitions displayed in Figs.\,\ref{fig:mmspecs}b) and c). To this end, we used, as initial distribution of Rydberg-atom--pair states the relative population in the two $\Omega=0$ and the $\Omega=1$ states that are accessed optically at the peak of the 32s$_{1/2}$33s$_{1/2}$ resonance, and consider transitions from these states to the pair states correlating to the 31p$_{3/2}$33s$_{1/2}$ and 32s$_{1/2}$33p$_{3/2}$ asymptotes that are accessible in a single-photon millimeter-wave transition. The relevant potential-energy functions are displayed in Fig.\,\ref{mmPEC}.

The attractive nature of the three potential functions associated with the 32s$_{1/2}$33s$_{1/2}$ asymptote implies, in a classical description of the nuclear motion, that the internuclear separation decreases with time. Because the potential curves of the states associated with the 31p$_{3/2}$33s$_{1/2}$ and 32s$_{1/2}$33p$_{3/2}$ asymptotes are less attractive than those of the states associated with the 32s$_{1/2}$33s$_{1/2}$ asymptote, the frequency of the 32s$_{1/2}$33s$_{1/2} \rightarrow  31$p$_{3/2}$33s$_{1/2}$ transition decreases, and that of the 32s$_{1/2}$33p$_{3/2} \leftarrow 32$s$_{1/2}$33s$_{1/2}$ increases, with time. This evolution (chirp) of the transition frequencies reduces the transition probabilities. The narrow bandwidth (1 MHz) of the millimeter-wave radiation implies that the main contribution to the millimeter-wave spectra originate from pairs with large internuclear separation $R$, for which the evolution (chirp) of the transition frequency is slow.

To model the intensity distribution, we introduce a cut-off chirp of 2 MHz/$\mu$s, beyond which we assume that the transitions are too weak and broad to be observed and determine the range of internuclear separations contributing to the millimeter-wave transition intensities accordingly. In this range, we calculate the distribution of transition frequencies from the potential-energy curves. For each frequency, we determine  the spectral intensities as sum over products of the initial populations in the $\Omega = 0$ and 1 states correlated to the 32s$_{1/2}$33s$_{1/2}$ asymptote and the relative transition probabilities
\begin{align}
P_\textrm{trans}(R) & =  \nonumber \\
& \sum_{i,f} f_{\Psi,i} (R) \overline{p_{\Psi,i}}(R) \left| \braket{\Psi_f(R) | \hat{\mu}_{0,\pm 1}   |  \Psi_i(R)} \right|^2 R^2 ,
\label{eq:mm_trans_prob}
\end{align}
where $f_{\Psi,i}(R)$ is the cutoff function described above, $\hat{\mu}_{0,\pm 1} $ is the sum of $\sigma^\pm$ and $\pi$ electric-dipole transition operators, and the double sum runs over all initially excited pair-state  $\Psi_i$ for a fixed value of $\Omega$ and all dipole-coupled final states $\Psi_f$.

The calculated spectra, convoluted with the experimental line-shape function of the corresponding atomic transitions, are displayed as dashed red lines in Figs.\,\ref{fig:mmspecs}b) and c) and are in qualitative agreement with the experimental ones. In particular, the shifts from the asymptotic transition frequencies, which correspond to the spectra of the isolated atoms displayed as lower traces, and the asymmetric line shapes (red-degraded for the 32s$_{1/2}33$s$_{1/2}\rightarrow 31$p$_{3/2}$33s$_{1/2}$ transition and blue degraded for the 32s$_{1/2}$33p$_{3/2} \leftarrow 32$s$_{1/2}$33s$_{1/2}$) are well described by the calculations. This agreement represents a further illustration of the ability of our potential model to describe the spectral and dynamical properties of Rydberg-atom pairs.

\section{Conclusions}

In this article we have presented a comprehensive survey of the Rydberg excitation spectrum in a dense (density $2\cdot 10^{11}$ atoms/cm$^{3}$), ultracold ($T=40$ $\mu$K) Cs gas recorded using an intense (peak intensity  $10^{8}$ W/cm$^2$) narrow-band (bandwidth 140 MHz) UV laser source.

Next to the strong lines corresponding to the atomic $n{\rm p}_{3/2} \leftarrow 6$s transitions, the spectra also exhibit a very rich structure of weaker, sharp spectral features, which are attributed to pairs of Rydberg atoms, also known as Rydberg macrodimers \cite{boisseau2002,samboy2011}, interacting through long-range electrostatic forces. Sharp features of three different origins have been identified: (i) lines with maxima at positions located very close to $n$s$_{1/2}(n+1)$s$_{1/2}$ and $n$s$_{1/2}(n-3)$f dissociation asymptotes and which correspond to the formation of weakly interacting Rydberg-atom pairs at long range, (ii) lines with maxima located above the $n$s$_{1/2}(n+1)$s$_{1/2}$ dissociation asymptotes and which correspond to excitation to flat potential regions near local maxima of the potentials of $\Omega=0$ and 1 molecular states correlated with $n$s$_{1/2}(n+1)$s$_{1/2}$ dissociation asymptotes, and (iii) "window" resonances located on the high-frequency side of the atomic transitions, which arise from avoided crossings between repulsive molecular potential curves correlated to $n$p$_{3/2}n$p$_{3/2}$ dissociation asymptotes and attractive potentials near $(n-4)$f$_{j}(n-3)$f$ _{j^\prime}$ ($j,j^\prime = 5/2,7/2$) dissociation asymptotes. The first category of resonances had been observed in earlier studies (see, e.g. Refs. \cite{farooqi2003,stanojevic2006,stanojevic2008,overstreet2007,overstreet2009}). The observation of the latter two categories (ii) and (iii) represents the first major result of our investigation.

The second major result is the demonstration, on the basis of a rigorous comparison of experimental and calculated spectra, that long-range molecular potential models provide a quantitatively correct and accurate description of the behavior of interacting Rydberg atoms in ultracold gases, provided that a sufficient number of channels and long-range interaction terms are considered. The potential model derived to describe Cs Rydberg-atom pairs adequately reproduces experimental UV-laser photoassociation spectra, millimeter-wave spectra of transitions between different Rydberg-atom--pair states, and the observed Penning-ionization dynamics of the Rydberg-atom pairs. Given the very high density of electronic states, it first appears astonishing that adiabatic potential models are at all adequate to describe the spectral and dynamical properties of interacting Rydberg-atom pairs. The reason for the usefulness of such models is that, in the range of interaction strengths and internuclear distances that are relevant for most experiments in ultracold Rydberg gases, the heavy nuclei are almost stationary on the time scale associated with the electronic motion, taken as the inverse state density, i.e., about 1 ns at $n\approx 24$.

In the discussion of dipole- and van der Waals-blockade effects in the excitation of Rydberg states, the attractive and repulsive potentials arising from the dipole-dipole interaction near the asymptotes of Rydberg-atom--pair states, are often represented, for clarity and simplicity, as two-level systems of the kind depicted in Fig. \ref{fig:mo}a. The present analysis of Rydberg-atom--pair states of Cs in the range of principal quantum number $n=22-36$, which includes all relevant potential curves and long-range interaction terms, demonstrates the limits of two-level Rydberg-blockade models. Indeed, many interacting molecular potentials give rise to a multitude of avoided crossings, resonances and level shifts. The range of observable phenomena is thus much wider than two-level blockade models predict, which on the one hand reduces their range of applicability, but on the other also offers new opportunities for scientific investigations and applications.

\begin{acknowledgments}
This work is supported financially by the Swiss National Science Foundation under Project Nr.~200020-159848, the NCCR QSIT, and the EU Initial Training Network COHERENCE under grant FP7-PEOPLE-2010-ITN-265031. We acknowledge the European Union H2020 FET Proactive project RySQ (grant N. 640378).
\end{acknowledgments}

\appendix*
\section{Calculation of matrix elements of $V'_{\rm int}$}

In this appendix, explicit expressions for the matrix elements of the atomic multipoles contributing to $V_\textrm{inter}$ (see Eq.~\eqref{eq:vint}) are derived in the product basis $\ket{n_\AR l_\AR j_\AR \omega_\AR , n_\BR l_\BR j_\BR \omega_\BR}$.

The matrix elements of an atomic multipole
\begin{align}
Q_{L,\omega}(\vec{r}) &= \sqrt{\frac{4 \pi}{2 L +1}} r^L Y_{L,\omega}(\hat{r})
\label{eq:qdefinition}
\end{align}
in this basis are given by
\begin{align}
&\bra{ n_i' l_i' j_i' \omega_i'} \hat{Q}_{L,\omega}(\vec{r}) \ket{n_i l_i j_i \omega_i} \nonumber \\
~~& = \sqrt{\frac{4 \pi}{2 L +1}} \bra{n_i' l_i' j_i'} \hat{r}^L \ket{n_i l_i j_i} \bra{l_i' j_i' \omega_i'} \hat{Y}_{L,\omega}(\hat{r}) \ket{l_i j_i \omega_i},
\end{align}
where the radial matrix elements
\begin{align}
& \bra{n_i' l_i' j_i'} \hat{r}^L \ket{n_i l_i j_i} \equiv {R_{n_i l_i j_i}^{n_i' l_i' j_i'}}^L \nonumber \\
~~& = \int \Psi^*_{n_i' l_i' j_i'}(r) r^L \Psi_{n_i l_i j_i}(r) dr
\end{align}
are evaluated numerically using the Numerov method~\cite{zimmerman1979} and a model core potential\,\cite{marinescu1994}. The angular matrix elements are evaluated using the analytical expressions\,\cite{varshalovich1988}
\begin{align}
&\bra{l_i' j_i' \omega_i'} \hat{Y}_{L,\omega}(\hat{r}) \ket{l_i j_i \omega_i} \nonumber \\
~~& = (-1)^{j_i + l_i' + L + \frac{1}{2}} \sqrt{(2 j_i + 1)(2 l_i + 1)}\nonumber \\
~~&~ \times \begin{Bmatrix}l_i & \frac{1}{2} & j_i \\ j_i' & L & l_i' \end{Bmatrix} \sqrt{\frac{2 L + 1}{4 \pi}}C_{l_i 0, L 0}^{l_i' 0} C_{j_i \omega_i, L \omega}^{j_i' \omega_i'} ~~,
\end{align}
where $C_{l_i \omega_i, L \omega}^{l_i' \omega_i'}$ are the Clebsch-Gordan coefficients and $\begin{Bmatrix}j_1 & j_2 & j_3 \\ j_4 & j_5 & j_6 \end{Bmatrix}$ is Wigner's 6-$j$ symbol.

As an example we give the full expression for the matrix elements of the lowest-order term in the multipole expansion, the dipole-dipole interaction:
\begin{align}
&\bra{n'_{\rm A} l'_{\rm A} j'_{\rm A} \omega'_{\rm A},n'_{\rm B} l'_{\rm B}j'_{\rm B}\omega'_{\rm B}}\hat{V}_{\rm dd}\ket{n_{\rm A} l_{\rm A}j_{\rm A}\omega_{\rm A},n_{\rm B} l_{\rm B}j_{\rm B}\omega_{\rm B}} \nonumber \\
~~&=(-1)^{j_{\rm A} + j_{\rm B} + l^\prime_{\rm A} + l^\prime_{\rm B}} R^{-3} {R_{n_{\rm A} l_{\rm A} j_{\rm A}}^{n_{\rm A}' l_{\rm A}' j_{\rm A}'}}^1 {R_{n_{\rm B} l_{\rm B} j_{\rm B}}^{n_{\rm B}' l_{\rm B}' j_{\rm B}'}}^1 \nonumber \\
~~&~\times \sqrt{(2 j_{\rm A} + 1) (2 j_{\rm B} + 1)(2 l_{\rm A} +1) ( 2 l_{\rm B} + 1)} \nonumber \\
~~&~\times {\Big(} C_{j_{\rm A} \omega_{\rm A}, 1 1}^{j_{\rm A}', \omega_{\rm A}'} C_{j_{\rm B} \omega_{\rm B}, 1 -1}^{j_{\rm B}', \omega_{\rm B}'} + \nonumber \\
~~&~ +2 C_{j_{\rm A} \omega_{\rm A}, 1 0}^{j_{\rm A}', \omega_{\rm A}'} C_{j_{\rm B} \omega_{\rm B}, 1 0}^{j_{\rm B}', \omega_{\rm B}'} +  C_{j_{\rm A} \omega_{\rm A}, 1 -1}^{j_{\rm A}', \omega_{\rm A}'} C_{j_{\rm B} \omega_{\rm B}, 1 1}^{j_{\rm B}', \omega_{\rm B}'} \Big) \nonumber \\
~~&~\times C_{l_{\rm A} 0, 1 0}^{l_{\rm A}' 0} C_{l_{\rm B} 0, 1 0}^{l_{\rm B}' 0} \begin{Bmatrix}l_{\rm A} & \frac{1}{2} & j_{\rm A} \\ j_{\rm A}' & L & l_{\rm A}' \end{Bmatrix} \begin{Bmatrix}l_{\rm B} & \frac{1}{2} & j_{\rm B} \\ j_{\rm B}' & L & l_{\rm B}' \end{Bmatrix} .
\label{eq:matrixelementsDD}
\end{align}

%\bibliography{pairstatespaper}

\end{document}